\documentstyle[11pt,bezier,epsf,amsfonts]{article}

\RequirePackage{times}
\RequirePackage{mathptm}

\begin{document}


\title{{\huge {\bf Embedding Quantum Universes into Classical Ones}}
\thanks{This paper has been completed
during the visits of the first author at the University of Technology
Vienna (1997) and of the third   author at the University of Auckland (1997).
The first   author has been
partially supported by AURC A18/XXXXX/62090/F3414056, 1996. The
second author was supported by DFG Research Grant No. HE 2489/2-1.}}
\author{{\large {\bf Cristian S.\  Calude},}\thanks{Computer Science Department,
The University of Auckland, Private Bag 92019, Auckland, New Zealand,
 e-mail: cristian@cs.auckland.ac.nz.} \quad
{\large {\bf Peter H.\ Hertling},}\thanks{Computer Science Department,
The University of Auckland, Private Bag 92019, Auckland, New Zealand,
e-mail: hertling@cs.auckland.ac.nz.}  \quad
{\large {\bf Karl  Svozil}}\thanks{Institut f\"ur Theoretische Physik,
University of Technology Vienna,
Wiedner Hauptstra\ss e 8-10/136,  A-1040 Vienna, Austria,
e-mail: svozil@tph.tuwien.ac.at.}}

\date{ }
\maketitle

\thispagestyle{empty}

\begin{abstract}
Do the partial order and ortholattice operations of a quantum logic correspond
to the logical implication and connectives of classical logic?  Re-phrased,
how far might a classical understanding of quantum mechanics
be, in principle, possible?  A celebrated result
by Kochen and Specker answers the above question in the negative.
However, this answer is just one among different possible ones, not all
negative.
It is our aim  to discuss the above question in terms of mappings of quantum
worlds into classical ones, more specifically, in terms of embeddings of
quantum logics
into classical logics; depending upon the type of restrictions imposed on
embeddings the question may get negative or positive answers.

\end{abstract}

\section{Introduction}

Quantum mechanics is a very successful theory which appears to
predict novel ``counterintuitive'' phenomena (see Wheeler \cite{wheeler},
Greenberger, Horne and Zeilinger \cite{green-horn-zei}) even
almost a century after
its development, cf.\ Schr{\"o}dinger \cite{schrodinger},
Jammer \cite{jammer:66,jammer1}.
Yet, it can be safely stated that quantum
theory is not understood (Feynman \cite{feynman-law}).
Indeed, it appears that progress is fostered by abandoning long--held
beliefs and concepts rather than by attempts to derive it from some
classical basis, cf.\ Greenberger and YaSin \cite{greenberger2},
Herzog, Kwiat, Weinfurter and Zeilinger \cite{hkwz}
and Bennett \cite{benn:94}.

But just how far might a classical understanding of quantum mechanics
be, in principle, possible?
We shall attempt an answer to this question in terms of mappings of quantum
worlds into
classical ones, more specifically, in terms of embeddings of quantum logics
into classical logics.

One physical motivation for this approach is a result proven for the
first time by  Kochen and Specker
\cite{kochen1} (cf. also Specker \cite{specker-60}, Zierler and
Schlessinger \cite{ZirlSchl-65} and John Bell \cite{bell-66}; see  reviews
by Mermin
\cite{mermin-93}, Svozil and Tkadlec \cite{svozil-tkadlec}, and a forthcoming
monograph by Svozil \cite{svozil-ql})
stating  the impossibility to ``complete'' quantum physics
by introducing noncontextual hidden parameter models.
Such a possible ``completion'' had been suggested, though in not very
concrete terms, by Einstein, Podolsky and Rosen (EPR) \cite{epr}. These authors
speculated that ``elements of physical reality'' exist irrespective of
whether they are actually observed. Moreover, EPR conjectured, the
quantum formalism can be ``completed'' or ``embedded'' into a larger
theoretical framework
which would reproduce the quantum theoretical results but would otherwise
be classical and deterministic from an algebraic and logical point of view.

A proper formalization of the term ``element of physical reality''
suggested by EPR can be given in terms of two-valued states or valuations,
which can take on only one of the two values $0$ and $1$, and which are
interpretable as the classical logical truth assignments {\it false} and
{\it true}, respectively.  Kochen and Specker's results
\cite{kochen1} state that for quantum systems representable by Hilbert
spaces of dimension higher than two, there does not exist any such valuation
$s: L\rightarrow \{0,1\}$ defined
on the set of closed linear subspaces of the space $L$
(these subspaces are interpretable as quantum mechanical propositions)
preserving the lattice operations and the orthocomplement,
even if one restricts the attention to lattice operations carried out among
commuting (orthogonal) elements.
As a consequence, the set of truth assignments on quantum logics is not
separating and not unital. That is, there exist  different quantum propositions
which cannot be distinguished by any classical truth assignment.

The Kochen and Specker result, as it is commonly argued, e.g.\ by
Peres \cite{peres} and Mermin \cite{mermin-93}, is directed against
the noncontextual hidden parameter program envisaged by EPR. Indeed, if one
takes into account the entire Hilbert logic  (of dimension
larger than two) and if one considers all states thereon, any truth
value assignment to quantum propositions prior to the actual measurement
yields a contradiction. This can be proven by finitistic means, that is,
with a finite number of one-dimensional closed linear subspaces (generating an
infinite  set whose intersection with the unit sphere is dense;
cf.\ Havlicek and Svozil \cite{havlicek}).
But, the Kochen--Specker argument continues, it is always possible to
prove the existence of separable valuations or truth assignments for
classical propositional systems identifiable with
Boolean algebras. Hence, there does not exist any  injective
morphism from a quantum logic into some Boolean algebra.

Since the previous reviews of the Kochen--Specker theorem by
Peres \cite{peres-91,peres}, Redhead \cite{redhead},
Clifton \cite{clifton-93}, Mermin \cite{mermin-93},
Svozil and Tkadlec \cite{svozil-tkadlec}, concentrated on
the nonexistence of classical noncontextual elements of physical reality,
we are going to discuss here some options and aspects of embeddings in
greater detail.
Particular emphasis will be given to embeddings of
quantum universes into classical ones which do not
necessarily preserve (binary lattice) operations identifiable with the
logical {\it or} and {\it and} operations. Stated pointedly, if one is
willing to abandon the preservation of quite commonly used logical
functions, then it is possible to give a classical meaning to quantum
physical statements, thus giving raise to an ``understanding'' of
quantum mechanics.

Quantum logic, according to
Birkhoff \cite{birkhoff-36}, Mackey \cite{ma-57},
Jauch \cite{jauch}, Kalmbach \cite{kalmbach-83},
Pulmannov{\'{a}} \cite{pulmannova-91},
identifies logical entities with Hilbert space entities.
In particular, elementary propositions $p,q,\ldots$  are associated
with closed linear subspaces of a Hilbert space through the origin
(zero vector); the implication relation $\leq$
is associated with the set theoretical subset relation $\subseteq$, and the
logical {\it or} $\vee$, {\it and} $\wedge$, and {\it not} $'$ operations
are associated with the set theoretic intersection $\cap$, with the
linear span $\oplus$ of subspaces and the orthogonal subspace $\perp$,
respectively.
The trivial logical
statement $1$ which is always true is identified with the entire Hilbert
space $H$, and its complement $\emptyset$ with the zero-dimensional
subspace (zero vector).
Two propositions $p$ and $q$ are orthogonal if
and only if $p\leq q'$.
Two propositions $p,q$ are co--measurable (commuting)
if and only if there exist mutually orthogonal propositions  $a,b,c$
such that $p=a\vee b$ and $q=a\vee c$.
Clearly, orthogonality implies co--measurability, since if $p$ and $q$
are orthogonal, we may identify
$a, b, c$ with $0,p,q$, respectively.
The negation of $p\leq q$ is denoted
by $p \not\leq q$.

\section{Varieties of embeddings}

One of the questions already raised in Specker's almost
forgotten first article \cite{specker-60}\footnote{In German.} concerned an
embedding of a quantum logical structure $L$ of propositions into a classical
universe represented by a Boolean algebra $B$. Thereby, it is taken as
a matter of principle that such an embedding should preserve as much
logico--algebraic structure as possible. An embedding of this kind can be
formalized as a mapping $\varphi :L\rightarrow B$  with the following
properties.\footnote{Specker had a modified notion of embedding in mind; see
below.} Let $p,q\in L$.

\begin{description}

\item[{\rm (i)}]
{\em Injectivity}:
two different quantum logical propositions are mapped into two
different propositions of the Boolean algebra, i.e., if $p\neq
q, $ then $ \varphi (p)\neq \varphi (q)$.

\item[{\rm (ii)}]
{\em Preservation of the order relation}:
if $p\leq q$,  then $\varphi (p) \leq \varphi (q)$.

\item[{\rm (iii)}]
{\em Preservation of ortholattice operations}, i.e.\ preservation of the
\begin{description}
\item[{\rm (ortho-)complement}:]
$\varphi(p')=\varphi (p)'$,
\item[{\it or} {\rm operation}:] $\varphi (p\vee q)=\varphi (p)
\vee \varphi (q)$,
\item[{\it and} {\rm operation}:] $\varphi (p\wedge q)=\varphi (p)
\wedge \varphi (q)$.
\end{description}
\end{description}

As it turns out,  we cannot have an embedding from the quantum universe
to the classical universe satisfying all three requirements (i)--(iii).
In particular, a head-on approach requiring (iii) is doomed to failure,
since the nonpreservation of ortholattice operations
among nonco--measurable propositions is quite evident, given the
nondistributive structure of quantum logics.

\subsection{Injective lattice morphisms}

\begin{figure}[htd]
\begin{center}
\unitlength 0.80mm
\linethickness{0.4pt}
\begin{picture}(125.00,60.73)
\put(60.00,0.00){\circle*{2.11}}
\put(30.00,30.00){\circle*{2.11}}
\put(60.00,59.67){\circle*{2.11}}
\put(90.00,30.00){\circle*{2.11}}
\put(60.00,0.00){\line(-1,1){30.00}}
\put(30.00,30.00){\line(1,1){30.00}}
\put(60.00,60.00){\line(1,-1){30.00}}
\put(90.00,30.00){\line(-1,-1){30.00}}
\put(65.00,0.00){\makebox(0,0)[lc]{$0=1'$}}
\put(65.00,60.00){\makebox(0,0)[lc]{$1=0'$}}
\put(30.00,25.00){\makebox(0,0)[cc]{$p_+$}}
\put(90.00,25.00){\makebox(0,0)[cc]{$q_-$}}
\put(60.00,0.00){\line(-5,3){50.00}}
\put(10.00,30.00){\line(5,3){50.00}}
\put(60.00,60.00){\line(5,-3){50.00}}
\put(110.00,30.00){\line(-5,-3){50.00}}
\put(10.00,30.00){\circle*{2.11}}
\put(10.00,25.00){\makebox(0,0)[cc]{$p_-$}}
\put(110.00,30.00){\circle*{2.11}}
\put(110.00,25.00){\makebox(0,0)[cc]{$q_+$}}
\end{picture}
\end{center}
\caption{\label{f-hd-mo2}
Hasse diagram of the ``Chinese lantern'' form of  $MO_2$.
\index{$MO_2$}
}
\end{figure}
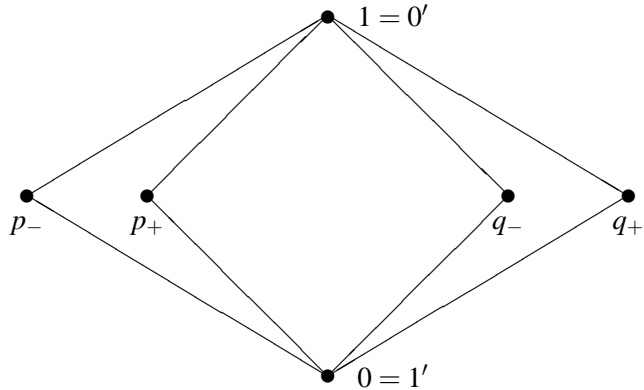

Here we shall review the rather evident fact that there does not
exist an injective lattice
morphism from any nondistributive lattice into a Boolean algebra.
We illustrate this obvious fact with an example that we need to
refer to later on in this paper;
the propositional structure encountered in the
quantum mechanics of spin state measurements of a spin one-half particle
along two different directions (mod~$\pi$), that is, the modular,
orthocomplemented lattice $MO_2$ drawn in Figure \ref{f-hd-mo2}
(where $p_-=(p_+)^\prime$ and $q_-=(q_+)^\prime$).

 Clearly, $MO_2$ is a nondistributive lattice, since for instance,
$$p_- \wedge (q_- \vee q_+)=p_- \wedge 1=p_-,$$ whereas
$$(p_- \wedge q_-)\vee (p_- \wedge q_+)=
0\vee0=0.$$ Hence,
$$p_- \wedge (q_- \vee q_+)\neq (p_- \wedge q_-)\vee (p_- \wedge
q_+).$$
In fact, $MO_2$ is the smallest orthocomplemented nondistributive
lattice.

The requirement (iii) that the embedding $\varphi$ preserves all
ortholattice operations (even for nonco--measurable and nonorthogonal
propositions) would
mean that
$\varphi(p_-) \wedge (\varphi(q_-) \vee \varphi (q_+))\neq (\varphi(p_-)
\wedge \varphi(q_-))\vee (\varphi(p_-) \wedge \varphi(q_+))$. That is, the
argument implies that
the distributive law is not satisfied in the range of $\varphi$.
But since the range of $\varphi$ is a subset of a Boolean algebra and for any
Boolean algebra the distributive law is satisfied,
this yields a  contradiction.

Could we still hope for a reasonable kind of embedding of a quantum
universe into a classical one by weakening our requirements, most notably
(iii)?
In the next three sections we are going to give different answers
to this question. In the first section we restrict the set of propositions
among which we wish to preserve the three operations {\em complement} $'$,
{\em or} $\vee$, and {\em and} $\wedge$. We will see that the
Kochen--Specker
result gives a very strong negative answer even when the restriction is
considerable. In the second section we analyze what happens if we try to
preserve not all operations but just the complement. Here we will obtain
a positive answer.
In the third section we discuss a different embedding which preserves the
order relation but no ortholattice operation.

\subsection{Injective order morphisms preserving ortholattice operations
among orthogonal propositions}

Let us follow Zierler and Schlessinger \cite{ZirlSchl-65} and Kochen and
Specker \cite{kochen1} and weaken (iii) by requiring that the
ortholattice operations need only to be preserved {\em among orthogonal}
propositions. As shown by Kochen and
Specker \cite{kochen1}, this is equivalent to the requirement of
separability by the  set of valuations or two-valued probability measures
or truth assignments on $L$.
As a matter of fact, Kochen and Specker \cite{kochen1} proved
nonseparability, but also much
more---the {\em nonexistence} of valuations on Hilbert lattices associated
with Hilbert spaces of dimension at least three. For related arguments
and conjectures, based upon a theorem by Gleason \cite{Gleason}, see
Zierler and Schlessinger \cite{ZirlSchl-65} and John Bell \cite{bell-66}.

Rather than rephrasing the Kochen and Specker argument
\cite{kochen1} concerning nonexistence
of valuations in three-dimensional Hilbert logics in its original form
or in terms of fewer
subspaces (cf.\ Peres \cite{peres}, Mermin \cite{mermin-93}), or of Greechie
diagrams, which represent orthogonality very nicely
(cf.\ Svozil and Tkadlec \cite{svozil-tkadlec},
Svozil \cite{svozil-ql}), we shall give two geometric
arguments which are derived from proof methods for Gleason's theorem
(see Piron \cite{piron-76}, Cooke, Keane, and Moran \cite{c-k-m},
and Kalmbach \cite{kalmbach-86}).

Let $L$ be the lattice of closed linear subspaces of the
three-dimensional real Hilbert space ${\Bbb R}^3$. A {\em two-valued
probability measure} or {\em valuation} on $L$ is a map
$v:L\to\{0,1\}$ which maps the zero-dimensional subspace
containing only the origin $(0,0,0)$ to $0$, the full space
${\Bbb R}^3$ to $1$, and which is additive on orthogonal subspaces.
This means that for two orthogonal subspaces $s_1, s_2 \in L$
the sum of the values $v(s_1)$ and $v(s_2)$ is equal to the
value of the linear span of $s_1$ and $s_2$. Hence,
if $s_1, s_2, s_3 \in L$ are a tripod of pairwise orthogonal
one-dimensional subspaces, then
\[ v(s_1) + v(s_2) + v(s_3) = v({\Bbb R}^3) = 1. \]
The valuation $v$ must map one of these subspaces to $1$ and the other
two to $0$.
We will show that there is {\it no}  such map.
In fact, we show that there is no map $v$ which is defined on all
one-dimensional subspaces of ${\Bbb R}^3$ and maps
{\em   exactly one subspace out of each tripod of pairwise
      orthogonal one-dimensional subspaces to $1$ and the other two to $0$}.

In the following two geometric proofs we often identify
a given one-dimensional subspace of ${\Bbb R}^3$ with one of its two
intersection points with the unit sphere
\[ S^2 = \{x \in {\Bbb R}^3 \ | \ ||x||=1\}\,. \]
In the statements  ``a point (on the unit sphere) has value $0$
(or value $1$)'' or
that ``two points (on the unit sphere) are orthogonal'' we always
mean the corresponding one-dimensional subspaces.
Note also that the intersection of a two-dimensional subspace with
the unit sphere is a great circle.

To start  the first proof,
let us assume that a function $v$ satisfying the above condition
exists.
Let us consider an arbitrary tripod of orthogonal points
and let us fix the point with value $1$. By a rotation we
can assume that it is the north pole with the coordinates
$(0,0,1)$. Then, by the condition above, all points
on the equator $\{(x,y,z) \in S^2\ | \ z=0\}$ must have value $0$
since they are orthogonal to the north pole.

\begin{figure}[htbp]
\centerline{ \epsfxsize12cm \epsfbox{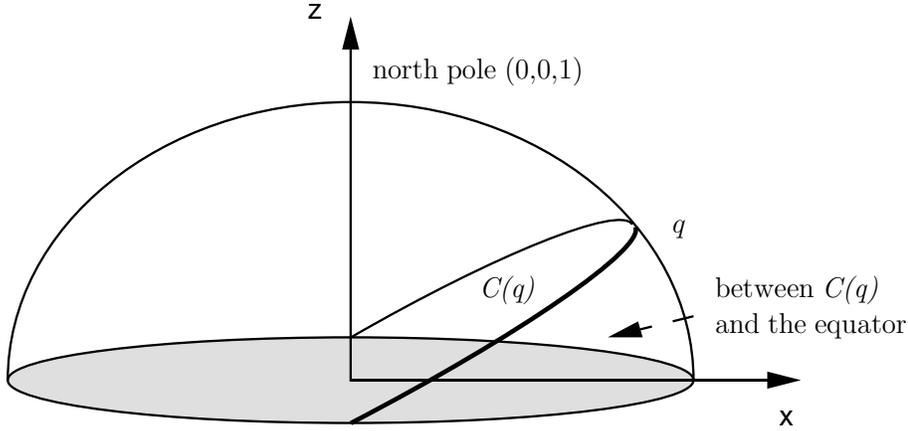} }
\vskip-2cm
\caption{The great circle $C(q)$.}
\label{figure:greatcircle}
\end{figure}

Let $q=(q_x,q_y,q_z)$ be a point in the northern
hemisphere, but not equal to the north pole, that is
$0< q_z < 1$. Let $C(q)$ be the unique
great circle which contains $q$ and the points
$\pm(q_y,-q_x,0)/\sqrt{q_x^2+q_y^2}$ in the equator, which are orthogonal
to $q$.
Obviously,  $q$ is the northern-most point on $C(q)$.
To see this, rotate the sphere around the $z$-axis
so that $q$ comes to lie in the $\{y=0\}$-plane;
see Figure \ref{figure:greatcircle}.
Then the two points in the equator orthogonal to $q$ are
just the points $\pm(0,1,0)$, and $C(q)$ is the intersection
of the plane through $q$ and $(0,1,0)$ with the unit sphere, hence
\[C(q) = \{p \in {\Bbb R}^3 \ | \ (\exists \ \alpha,\beta \in {\Bbb R}) \
     \alpha^2 + \beta^2 =1 \ \mbox{\rm and } p=\alpha q + \beta
     (0,1,0) \}\,.\]
This shows that $q$ has the largest $z$-coordinate among all
points in $C(q)$.

Assume that $q$ has value $0$. We claim that then all points
on $C(q)$ must have value $0$. Indeed, since $q$ has value
$0$ and the orthogonal point $(q_y,-q_x,0)/\sqrt{q_x^2+q_y^2}$
on the equator also has value $0$, the one-dimensional subspace
orthogonal to both of them must have value $1$.
But this subspace is orthogonal to all points on $C(q)$.
Hence all points on $C(q)$ must have value $0$.

Now we can apply the same argument to any point $\tilde{q}$ on
$C(q)$ (by the last consideration $\tilde{q}$ must have value $0$)
and derive that all points on $C(\tilde{q})$ have value $0$.
The great circle $C(q)$ divides the northern hemisphere into two
regions, one containing the north pole,
the other consisting of the points below $C(q)$ or
``lying between $C(q)$ and the equator'',
see Figure \ref{figure:greatcircle}.
The circles $C(\tilde{q})$ with $\tilde{q} \in C(q)$
certainly cover the region between $C(q)$ and the equator.\footnote{This will
be shown formally in the proof of the geometric lemma below.}
Hence any point in this region must have value $0$.

But the circles $C(\tilde{q})$ cover also a part
of the other region. In fact, we can iterate this process.
We say that a point $p$ in the northern hemisphere
{\em can be reached} from a point $q$ in the
northern hemisphere, if there is a finite
sequence of points $q=q_0, q_1, \ldots, q_{n-1}, q_n=p$
in the northern hemisphere such that $q_i\in C(q_{i-1})$
for $i=1,\ldots,n$.
Our analysis above shows that if $q$ has value $0$ and
$p$ can be reached from $q$, then also $p$ has value $0$.

The following geometric lemma due to Piron \cite{piron-76}
(see also
Cooke, Keane, and Moran \cite{c-k-m} or Kalmbach \cite{kalmbach-86})
is a consequence of the fact that the curve $C(q)$ is tangent to
the horizontal plane through the point $q$:
\begin{quote}
{\it If $q$ and $p$ are points in the northern hemisphere
with $p_z < q_z$, then $p$ can be reached from $q$.}
\end{quote}
This result will be proved in Appendix A.
We conclude that, if a point $q$ in the northern hemisphere has value $0$,
then every point $p$ in the northern hemisphere with $p_z < q_z$ must
have value $0$ as well.

Consider the tripod $(1,0,0), (0,{1 \over \sqrt{2}},{1 \over \sqrt{2}}),
(0,-{1 \over \sqrt{2}},{1 \over \sqrt{2}})$. Since $(1,0,0)$ (on the equator)
has value $0$,
one of the two other points has value $0$ and one has
value $1$. By the geometric lemma and our above considerations
this implies that all points $p$ in the northern
hemisphere with $p_z<{ 1\over\sqrt{2}}$ must have value $0$
and all points $p$ with $p_z>{1\over\sqrt{2}}$ must have value $1$.
But now we can choose any point $p^\prime$ with
${1\over\sqrt{2}} < p^\prime_z < 1$
as our new north pole and  deduce that the valuation
must have the same form with respect to this
pole. This is clearly impossible.
Hence, we have proved our assertion that there is no mapping on the
set of all one-dimensional subspaces of ${\Bbb R}^3$
which maps one space out of each tripod of pairwise orthogonal
one-dimensional subspaces to $1$ and the other two to $0$.

In the following we give a second topological and
geometric proof for this fact.
In this proof we shall not use the geometric lemma above.

Fix an arbitrary point on the unit sphere with value $0$.
The  great circle consisting of points orthogonal to this
point splits into two disjoint sets, the set of points
with value $1$, and the set of points orthogonal to these
points. They have value $0$. If one of these two
sets were open, then the other had to be open as well.
But this is impossible since the circle is connected
and cannot be the union of two disjoint open sets.
Hence the circle must contain a point $p$ with value $1$
and a sequence of points $q(n)$, $n=1,2,\ldots$ with value $0$ converging
to $p$. By a rotation we can assume that $p$ is the
north pole and the circle lies in the $\{y=0\}$-plane.
Furthermore we can assume that all points $q_n$ have
the same sign in the $x$-coordinate. Otherwise,  choose
an infinite subsequence of the sequence $q(n)$ with this property.
In fact, by a rotation we can assume that all points $q(n)$ have
positive $x$-coordinate (i.e.\ all points $q(n)$, $n=1,2,\ldots$
lie as the point $q$ in  Figure \ref{figure:greatcircle} and approach the
north pole as $n$ tends to infinity).
All points on the equator have value $0$.
By the first step in the proof of the geometric lemma in the appendix,
all points in the northern hemisphere which lie between $C(q(n))$
(the great circle through $q(n)$ and $\pm(0,1,0)$)
and the equator can be reached from $q(n)$. Hence, as we
have seen in the first proof, $v(q(n))=0$ implies that
all these points must have value $0$.
Since $q(n)$ approaches
the north pole, the union of the regions between $C(q(n))$ and
the equator is equal to the open right half
$\{q \in S^2\ | \ q_z>0, q_x>0\}$
of the northern hemisphere.
%
%
 Hence all points in this set
have value $0$.  Let $q$ be a point in the left half
$\{q \in S^2\ | \ q_z>0, q_x<0\}$ of the northern hemisphere.
It forms a tripod together with the point
$(q_y,-q_x,0)/\sqrt{q_x^2+q_y^2}$ in the equator and the point
$(-q_x,-q_y,{q_x^2+q_y^2 \over q_z}) /
        ||(-q_x,-q_y,{q_x^2+q_y^2 \over q_z})||$
in the right half. Since these two points have value $0$,
the point $q$ must have value $1$. Hence all points
in the left half of the northern hemisphere must have
value $1$. But this leads to a contradiction because
there are tripods with two points in the left half,
for example the tripod
$(-{1 \over 2},{1 \over \sqrt{2}},{1 \over 2})$,
$(-{1 \over 2},-{1 \over \sqrt{2}},{1 \over 2})$,
$({1 \over \sqrt{2}},0,{1 \over \sqrt{2}})$.
This ends the second proof for the fact that
there is no two-valued probability measure
on the lattice of subspaces of the three-dimensional
Euclidean space which preserves the ortholattice operations
at least for orthogonal elements.

\subsection{Injective morphisms preserving order as well as {\it
or} and
{\it and} operations}
\label{section:2.3a}

We have seen that we cannot hope to preserve the ortholattice operations,
not even when we restrict ourselves to  operations
among orthogonal propositions.

An even stronger weakening of condition (iii) would be to require
preservation of
ortholattice operations merely among the center $C$,  i.e., among those
propositions which are
co--measurable (commuting) with all other propositions. It is not difficult
to prove that in the case
of complete Hilbert lattices (and not mere subalgebras thereof), the
center consists of just the least lower and the greatest upper bound
$C=\{0,1\}$ and thus is
isomorphic to the two-element Boolean algebra ${\bf 2}=\{0,1\}$.
As it turns out,  the requirement is trivially fulfilled and its
implications are quite trivial as well.

Another weakening of (iii) is to restrict oneself to particular physical
states and study the embeddability of quantum logics under these
constraints; see Bell, Clifton \cite{bell-clifton}.

In the following sections we analyze a completely different option: Is
it possible
to embed quantum logic into a Boolean algebra when one does not demand
preservation of all ortholattice operations?

One method of embedding an arbitrary partially ordered set
into
a concrete orthomodular lattice which in turn can be embedded into a
Boolean algebra has been used by Kalmbach \cite{kalmbach-77} and
extended by Harding \cite{harding-91} and Mayet and Navara
\cite{navara-95}. In
these {\it Kalmbach embeddings}, as they may be called,
the meets and joins are preserved but not the complement.

The Kalmbach embedding of some bounded lattice $L$ into a concrete
orthomodular lattice
$K(L)$ may be thought of as the pasting of Boolean algebras
corresponding to all maximal chains of $L$ \cite{harding-priv}.

First, let us consider  linear chains
$0=a_0\rightarrow
a_1\rightarrow a_2\rightarrow \cdots \rightarrow 1=a_m$.
Such chains
generate Boolean algebras ${\bf 2}^{m-1}$ in the following way: from the first
nonzero element $a_1$ on to the greatest element $1$, form
$A_n=a_n\wedge (a_{n-1})'$, where $(a_{n-1})'$ is the complement of $a_{n-1}$
relative to $1$; i.e., $(a_{n-1})'=1-a_{n-1}$.  $A_n$ is then an atom of the
Boolean algebra generated by the bounded chain
$0=a_0\rightarrow
a_1\rightarrow a_2\rightarrow \cdots \rightarrow 1$.

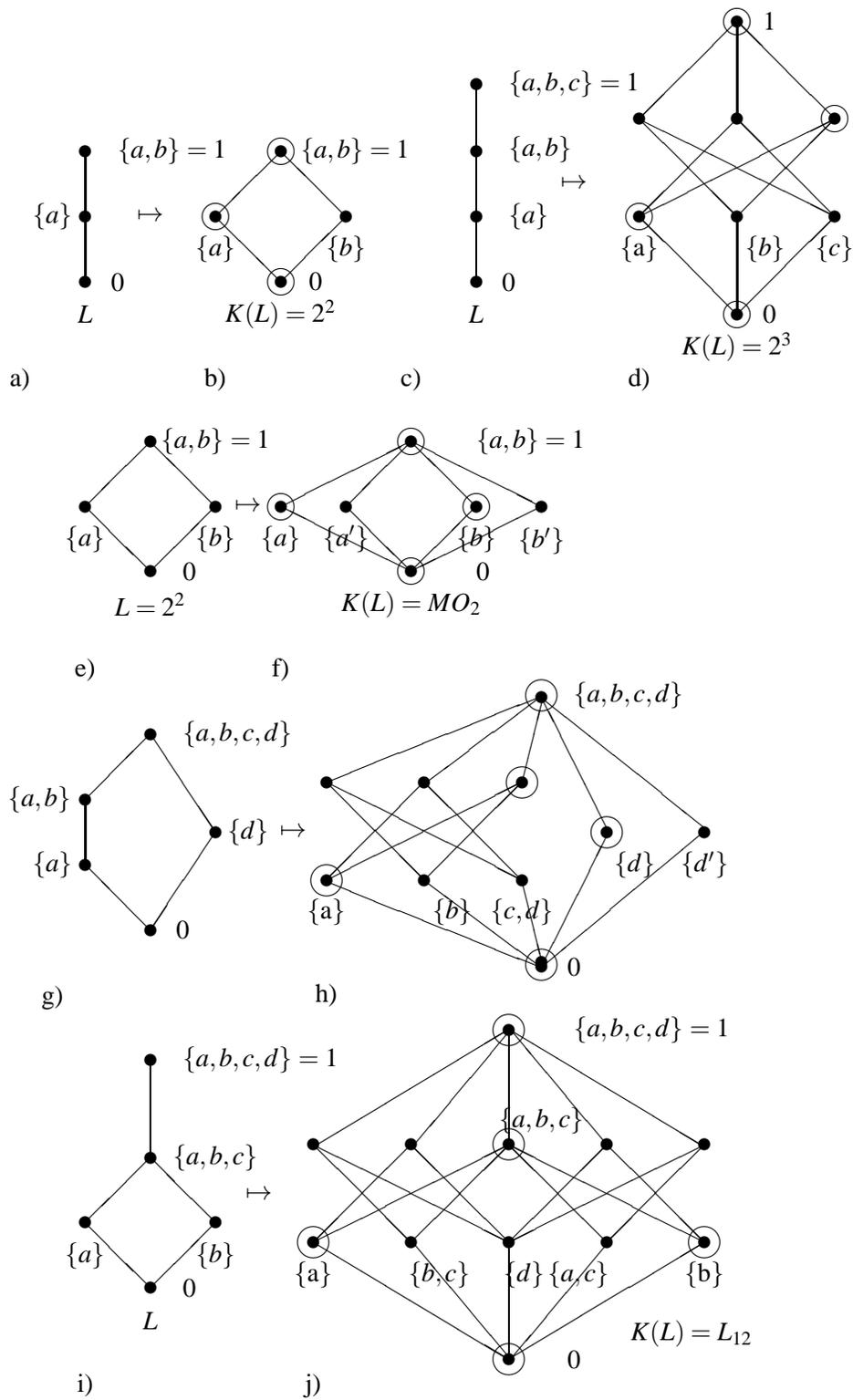
\begin{figure}
\begin{center}
\unitlength 0.95mm
\linethickness{0.4pt}
\begin{picture}(127.00,211.52)
\put(10.00,169.52){\circle*{1.89}}
\put(10.00,169.52){\line(0,1){10.00}}
\put(10.00,179.52){\circle*{1.89}}
\put(10.00,189.52){\circle*{1.89}}
\put(10.00,179.52){\line(0,1){10.00}}
\put(15.00,169.52){\makebox(0,0)[cc]{0}}
\put(5.00,179.52){\makebox(0,0)[cc]{$\{a\}$}}
\put(15.00,189.52){\makebox(0,0)[lc]{$\{a,b\}=1$}}
\put(40.00,169.52){\circle*{1.89}}
\put(40.00,189.52){\circle*{1.89}}
\put(30.00,179.52){\circle*{1.89}}
\put(50.00,179.52){\circle*{1.89}}
\put(50.00,179.52){\line(-1,1){10.00}}
\put(40.00,189.52){\line(-1,-1){10.00}}
\put(30.00,179.52){\line(1,-1){10.00}}
\put(40.00,169.52){\line(1,1){10.00}}
\put(45.33,169.52){\makebox(0,0)[cc]{$0$}}
\put(51.43,189.52){\makebox(0,0)[cc]{$\{a,b\}=1$}}
\put(0.00,154.52){\makebox(0,0)[cc]{a)}}
\put(30.00,154.52){\makebox(0,0)[cc]{b)}}
\put(70.00,169.52){\circle*{1.89}}
\put(70.00,169.52){\line(0,1){10.00}}
\put(70.00,179.52){\circle*{1.89}}
\put(70.00,189.52){\circle*{1.89}}
\put(70.00,179.52){\line(0,1){10.00}}
\put(75.00,169.52){\makebox(0,0)[cc]{0}}
\put(75.00,179.52){\makebox(0,0)[lc]{$\{a\}$}}
\put(60.00,154.52){\makebox(0,0)[cc]{c)}}
\put(70.00,199.85){\circle*{1.89}}
\put(70.00,189.85){\line(0,1){10.00}}
\put(75.00,189.85){\makebox(0,0)[lc]{$\{a,b\}$}}
\put(75.00,199.85){\makebox(0,0)[lc]{$\{a,b,c\}=1$}}
\put(20.00,179.52){\makebox(0,0)[cc]{$\mapsto$}}
\put(85.00,184.52){\makebox(0,0)[cc]{$\mapsto$}}
\put(47.00,77.67){\circle*{1.89}}
\put(62.00,77.67){\circle*{1.89}}
\put(77.00,77.67){\circle*{1.89}}
\put(47.00,92.67){\circle*{1.89}}
\put(62.00,92.67){\circle*{1.89}}
\put(77.00,92.67){\circle*{1.89}}
\put(110.00,209.52){\circle*{1.89}}
\put(110.00,164.52){\circle*{1.89}}
\put(110.00,164.52){\line(1,1){15.00}}
\put(77.00,77.67){\line(-1,1){15.00}}
\put(110.00,194.52){\line(0,1){15.00}}
\put(110.00,209.52){\line(-1,-1){15.00}}
\put(47.00,92.67){\line(1,-1){15.00}}
\put(110.00,179.52){\line(0,-1){15.00}}
\put(110.00,164.52){\line(-1,1){15.00}}
\put(47.00,77.67){\line(2,1){30.00}}
\put(125.00,194.52){\line(-1,1){15.00}}
\put(62.00,92.67){\line(-1,-1){15.00}}
\put(47.00,92.67){\line(2,-1){30.00}}
\put(62.00,77.67){\line(1,1){15.00}}
\put(47.00,72.67){\makebox(0,0)[cc]{\{a\}}}
\put(66.33,72.67){\makebox(0,0)[cc]{$\{b\}$}}
\put(77.00,72.67){\makebox(0,0)[cc]{$\{c,d\}$}}
\put(115.00,164.52){\makebox(0,0)[cc]{$0$}}
\put(115.00,209.52){\makebox(0,0)[cc]{$1$}}
\put(95.00,154.52){\makebox(0,0)[cc]{d)}}
\put(110.00,159.52){\makebox(0,0)[cc]{$K(L)=2^3$}}
\put(20.00,125.00){\circle*{1.89}}
\put(20.00,145.00){\circle*{1.89}}
\put(10.00,135.00){\circle*{1.89}}
\put(30.00,135.00){\circle*{1.89}}
\put(30.00,135.00){\line(-1,1){10.00}}
\put(20.00,145.00){\line(-1,-1){10.00}}
\put(10.00,135.00){\line(1,-1){10.00}}
\put(20.00,125.00){\line(1,1){10.00}}
\put(25.00,125.00){\makebox(0,0)[lc]{0}}
\put(30.00,145.00){\makebox(0,0)[cc]{$\{a,b\}=1$}}
\put(20.00,120.00){\makebox(0,0)[cc]{$L=2^2$}}
\put(10.00,110.00){\makebox(0,0)[cc]{e)}}
\put(30.00,174.52){\makebox(0,0)[cc]{$\{a\}$}}
\put(50.00,174.52){\makebox(0,0)[cc]{$\{b\}$}}
\put(10.00,130.00){\makebox(0,0)[cc]{$\{a\}$}}
\put(30.00,130.00){\makebox(0,0)[cc]{$\{b\}$}}
\put(60.00,125.00){\circle*{1.89}}
\put(60.00,145.00){\circle*{1.89}}
\put(50.00,135.00){\circle*{1.89}}
\put(70.00,135.00){\circle*{1.89}}
\put(70.00,135.00){\line(-1,1){10.00}}
\put(60.00,145.00){\line(-1,-1){10.00}}
\put(50.00,135.00){\line(1,-1){10.00}}
\put(60.00,125.00){\line(1,1){10.00}}
\put(70.00,125.00){\makebox(0,0)[lc]{0}}
\put(70.00,145.00){\makebox(0,0)[lc]{$\{a,b\}=1$}}
\put(50.00,130.00){\makebox(0,0)[cc]{$\{a'\}$}}
\put(70.00,130.00){\makebox(0,0)[cc]{$\{b\}$}}
\put(40.00,135.00){\circle*{1.89}}
\put(80.00,135.00){\circle*{1.89}}
\put(60.00,120.00){\makebox(0,0)[cc]{$K(L)=MO_2$}}
\put(40.00,110.00){\makebox(0,0)[cc]{f)}}
\put(60.00,125.00){\line(-2,1){20.00}}
\put(40.00,135.00){\line(2,1){20.00}}
\put(60.00,145.00){\line(2,-1){20.00}}
\put(80.00,135.00){\line(-2,-1){20.00}}
\put(40.00,130.00){\makebox(0,0)[cc]{$\{a\}$}}
\put(80.00,129.67){\makebox(0,0)[cc]{$\{b'\}$}}
\put(35.00,135.00){\makebox(0,0)[cc]{$\mapsto$}}
\put(10.00,164.52){\makebox(0,0)[cc]{$L$}}
\put(40.00,164.52){\makebox(0,0)[cc]{$K(L)=2^2$}}
\put(70.00,164.52){\makebox(0,0)[cc]{$L$}}
\put(20.00,100.00){\circle*{1.89}}
\put(20.00,70.00){\circle*{1.89}}
\put(10.00,80.00){\circle*{1.89}}
\put(10.00,90.00){\circle*{1.89}}
\put(30.00,85.00){\circle*{1.89}}
\put(20.00,70.00){\line(-1,1){10.00}}
\put(10.00,80.00){\line(0,1){10.00}}
\put(10.00,90.00){\line(1,1){10.00}}
\put(20.00,100.00){\line(2,-3){10.00}}
\put(30.00,85.00){\line(-2,-3){10.00}}
\put(3.00,90.00){\makebox(0,0)[cc]{$\{a,b\}$}}
\put(5.00,80.00){\makebox(0,0)[cc]{$\{a\}$}}
\put(35.00,85.00){\makebox(0,0)[cc]{$\{d\}$}}
\put(25.00,100.00){\makebox(0,0)[lc]{$\{a,b,c,d\}$}}
\put(25.00,70.00){\makebox(0,0)[cc]{$0$}}
\put(95.00,179.52){\circle*{1.89}}
\put(110.00,179.52){\circle*{1.89}}
\put(125.00,179.52){\circle*{1.89}}
\put(95.00,194.52){\circle*{1.89}}
\put(110.00,194.52){\circle*{1.89}}
\put(125.00,194.52){\circle*{1.89}}
\put(125.00,179.52){\line(-1,1){15.00}}
\put(95.00,194.52){\line(1,-1){15.00}}
\put(95.00,179.52){\line(2,1){30.00}}
\put(110.00,194.52){\line(-1,-1){15.00}}
\put(95.00,194.52){\line(2,-1){30.00}}
\put(110.00,179.52){\line(1,1){15.00}}
\put(95.00,174.52){\makebox(0,0)[cc]{\{a\}}}
\put(114.33,174.52){\makebox(0,0)[cc]{$\{b\}$}}
\put(125.00,174.52){\makebox(0,0)[cc]{$\{c\}$}}
\put(90.00,85.00){\circle*{1.89}}
\put(105.00,85.00){\circle*{1.89}}
\put(94.33,80.00){\makebox(0,0)[cc]{$\{d\}$}}
\put(105.00,80.00){\makebox(0,0)[cc]{$\{d'\}$}}
\put(80.00,105.67){\circle*{1.89}}
\put(80.00,65.00){\circle*{1.89}}
\put(47.00,92.67){\line(5,2){33.00}}
\put(80.00,105.87){\line(-4,-3){18.00}}
\put(77.00,92.67){\line(1,4){3.33}}
\put(80.00,64.33){\circle*{1.89}}
\put(47.00,77.33){\line(5,-2){33.00}}
\put(80.00,64.13){\line(-4,3){18.00}}
\put(77.00,77.33){\line(1,-4){3.33}}
\put(80.00,64.33){\line(1,2){10.33}}
\put(90.33,85.00){\line(-1,2){10.33}}
\put(80.00,105.67){\line(5,-4){25.33}}
\put(105.33,85.40){\line(-6,-5){25.33}}
\put(85.00,64.33){\makebox(0,0)[cc]{$0$}}
\put(85.00,106.00){\makebox(0,0)[lc]{$\{a,b,c,d\}$}}
\put(80.00,64.67){\circle{4.85}}
\put(47.00,77.67){\circle{4.85}}
\put(77.00,92.67){\circle{4.85}}
\put(80.00,106.00){\circle{4.85}}
\put(90.00,85.00){\circle{4.85}}
\put(20.00,50.00){\circle*{1.89}}
\put(20.00,15.00){\circle*{1.89}}
\put(20.00,35.00){\circle*{1.89}}
\put(10.00,25.00){\circle*{1.89}}
\put(30.00,25.00){\circle*{1.89}}
\put(30.00,25.00){\line(-1,1){10.00}}
\put(20.00,35.00){\line(-1,-1){10.00}}
\put(10.00,25.00){\line(1,-1){10.00}}
\put(20.00,15.00){\line(1,1){10.00}}
\put(25.00,15.00){\makebox(0,0)[lc]{0}}
\put(30.00,35.00){\makebox(0,0)[cc]{$\{a,b,c\}$}}
\put(20.00,10.00){\makebox(0,0)[cc]{$L$}}
\put(10.00,0.00){\makebox(0,0)[cc]{i)}}
\put(10.00,20.00){\makebox(0,0)[cc]{$\{a\}$}}
\put(30.00,20.00){\makebox(0,0)[cc]{$\{b\}$}}
\put(36.67,29.33){\makebox(0,0)[cc]{$\mapsto$}}
\put(5.00,59.67){\makebox(0,0)[cc]{g)}}
\put(47.00,60.00){\makebox(0,0)[cc]{h)}}
\put(20.00,35.00){\line(0,1){15.00}}
\put(25.00,50.00){\makebox(0,0)[lc]{$\{a,b,c,d\}=1$}}
\put(45.00,22.00){\circle*{1.89}}
\put(60.00,22.00){\circle*{1.89}}
\put(75.00,22.00){\circle*{1.89}}
\put(45.00,37.00){\circle*{1.89}}
\put(60.00,37.00){\circle*{1.89}}
\put(75.00,37.00){\circle*{1.89}}
\put(75.00,22.00){\line(-1,1){15.00}}
\put(45.00,37.00){\line(1,-1){15.00}}
\put(45.00,22.00){\line(2,1){30.00}}
\put(60.00,37.00){\line(-1,-1){15.00}}
\put(45.00,37.00){\line(2,-1){30.00}}
\put(60.00,22.00){\line(1,1){15.00}}
\put(45.00,17.00){\makebox(0,0)[cc]{\{a\}}}
\put(64.33,17.00){\makebox(0,0)[cc]{$\{b,c\}$}}
\put(77.33,17.00){\makebox(0,0)[cc]{$\{d\}$}}
\put(45.00,22.00){\circle{4.85}}
\put(75.00,37.00){\circle{4.85}}
\put(105.00,22.00){\circle*{1.89}}
\put(90.00,22.00){\circle*{1.89}}
\put(105.00,37.00){\circle*{1.89}}
\put(90.00,37.00){\circle*{1.89}}
\put(75.00,22.00){\line(1,1){15.00}}
\put(105.00,37.00){\line(-1,-1){15.00}}
\put(105.00,22.00){\line(-2,1){30.00}}
\put(90.00,37.00){\line(1,-1){15.00}}
\put(105.00,37.00){\line(-2,-1){30.00}}
\put(90.00,22.00){\line(-1,1){15.00}}
\put(105.00,17.00){\makebox(0,0)[cc]{\{b\}}}
\put(85.67,17.00){\makebox(0,0)[cc]{$\{a,c\}$}}
\put(105.00,22.00){\circle{4.85}}
\put(45.00,36.67){\line(5,3){30.00}}
\put(75.00,54.67){\line(5,-3){30.00}}
\put(90.00,37.00){\line(-5,6){14.72}}
\put(75.28,54.67){\line(-5,-6){14.72}}
\put(75.00,37.00){\line(0,1){17.33}}
\put(75.00,54.67){\circle*{1.89}}
\put(75.00,54.67){\circle{4.85}}
\put(45.00,22.00){\line(5,-3){30.00}}
\put(75.00,4.00){\line(5,3){30.00}}
\put(90.00,21.67){\line(-5,-6){14.72}}
\put(75.28,4.00){\line(-5,6){14.72}}
\put(75.00,21.67){\line(0,-1){17.33}}
\put(75.00,4.00){\circle*{1.89}}
\put(75.00,4.00){\circle{4.85}}
\put(45.00,0.00){\makebox(0,0)[cc]{j)}}
\put(85.00,4.00){\makebox(0,0)[cc]{$0$}}
\put(85.00,54.67){\makebox(0,0)[lc]{$\{a,b,c,d\}=1$}}
\put(80.00,40.67){\makebox(0,0)[cc]{$\{a,b,c\}$}}
\put(40.00,169.52){\circle{4.00}}
\put(30.00,179.52){\circle{4.00}}
\put(40.00,189.52){\circle{4.00}}
\put(110.00,164.52){\circle{4.00}}
\put(95.00,179.52){\circle{4.00}}
\put(125.00,194.52){\circle{4.00}}
\put(110.00,209.52){\circle{4.00}}
\put(60.00,145.00){\circle{4.00}}
\put(40.00,135.00){\circle{4.00}}
\put(70.00,135.00){\circle{4.00}}
\put(60.00,125.00){\circle{4.00}}
\put(103.00,8.00){\makebox(0,0)[cc]{$K(L)=L_{12}$}}
\put(41.90,85.00){\makebox(0,0)[cc]{$\mapsto$}}
\end{picture}
\end{center}
\caption{\label{f-thech}
Examples of Kalmbach embeddings.}
\end{figure}

Take, for example, a three-element chain
$0= a_0\rightarrow \{a\}\equiv a_1\rightarrow
\{a,b\}\equiv 1=a_2$
as depicted in Figure
\ref{f-thech}a).
In this case,
\begin{eqnarray*}
A_1&=&a_1\wedge (a_0)'=a_1\wedge 1\equiv \{a\}\wedge \{a,b\}=\{a\},\\
A_2&=&a_2\wedge (a_1)'=1\wedge (a_1)'\equiv \{a,b\}\wedge \{b\}=\{b\}.
\end{eqnarray*}
This construction results in a four-element Boolean Kalmbach lattice
$K(L)={\bf 2}^2$ with the two atoms $\{a\}$ and $\{b\}$
given in Figure \ref{f-thech}b).

Take, as a second example, a four-element chain
$0= a_0\rightarrow \{a\}\equiv a_1\rightarrow \{a,b\}
\rightarrow \{a,b,c\}\equiv 1=a_3$
as depicted in Figure
\ref{f-thech}c).
In this case,
\begin{eqnarray*}
A_1&=&a_1\wedge (a_0)'=a_1\wedge 1\equiv \{a\}\wedge \{a,b,c\}=\{a\},\\
A_2&=&a_2\wedge (a_1)'\equiv \{a,b\}\wedge \{b,c\}=\{b\},\\
A_3&=&a_3\wedge (a_2)'=1\wedge (a_2)'\equiv \{a,b,c\}\wedge \{c\}=\{c\}.\\
\end{eqnarray*}
This construction results in an eight-element Boolean Kalmbach lattice
$K(L)={\bf 2}^3$ with the three atoms $\{a\}$, $\{b\}$ and $\{c\}$
depicted in Figure \ref{f-thech}d).

To apply Kalmbach's
construction to any bounded lattice, all Boolean
algebras generated by the maximal chains of the lattice are pasted
together.
An element common to
two or more maximal chains must be common to the blocks they generate.

Take, as a third example, the Boolean lattice ${\bf 2}^2$ drawn in Figure
\ref{f-thech}e).
${\bf 2}^2$ contains two
linear chains of length three which are pasted together horizontally at
their smallest and biggest elements.  The resulting
Kalmbach lattice $K({\bf 2}^2)=MO_2$ is of the ``Chinese lantern'' type,
see Figure \ref{f-thech}f).

Take, as a fourth example, the pentagon drawn in Figure
\ref{f-thech}g).
It contains two
linear chains: one is of length three, the other is of length 4. The
resulting Boolean algebras ${\bf 2}^2$ and ${\bf 2}^3$ are again horizontally
pasted together at their extremities $0,1$.
The resulting
Kalmbach lattice is
given in Figure
\ref{f-thech}h).

In the
fifth example drawn in Figure
\ref{f-thech}i), the lattice has two maximal chains
which
share a common element.  This element is common to the two Boolean
algebras, hence central in $K(L)$.
The construction of the five atoms proceeds as follows:
\begin{eqnarray*}
A_1&=& \{a\}\wedge \{a,b,c,d\}=\{a\},\\
A_2&=& \{a,b,c\}\wedge \{b,c,d\}=\{b,c\},\\
A_3&=&B_3= \{a,b,c,d\}\wedge \{d\}=\{d\},\\
B_1&=& \{b\}\wedge \{a,b,c,d\}=\{b\},\\
B_2&=& \{a,b,c\}\wedge \{a,c,d\}=\{a,c\},\\
\end{eqnarray*}
where the two sets of atoms
$\{A_1,A_2,A_3=B_3\}$ and
$\{B_1,B_2,B_3=A_3\}$ span two Boolean algebras ${\bf 2}^3$ pasted
together at the extremities and at $A_3=B_3$ and $A_3'=B_3'$.
The resulting lattice is ${\bf 2}\times MO_2=L_{12}$ depicted in
Figure \ref{f-thech}j).

\subsection{Injective morphisms preserving order and complementation}
\label{section:2.3}

In the following, we shall show that {\em any orthoposet
can be embedded into a Boolean algebra} where in this case
by an {\em embedding} we understand  an {\em injective mapping
preserving the order relation and the orthocomplementation}.

A slightly stronger version of this fact using
more topological notions has already been shown by
Katrno\v{s}ka \cite{katrnoska-82}.
Zierler and Schlessinger constructed embeddings
with more properties for
orthomodular orthoposets \cite[Theorem 2.1]{ZirlSchl-65}
and mentioned another slightly
stronger version of the result above without explicit proof
\cite[Section 2, Remark 2]{ZirlSchl-65}.

For completeness sake we give the precise definition
of an orthoposet.
An {\em orthoposet} (or {\em orthocomplemented poset})
$(L,\leq,0,1,')$ is a set $L$ which is
endowed with a partial ordering
$\leq$, (i.e.\ a subset $\leq$ of $L \times L$ satisfying
(1) $p \leq p$, (2) if $p \leq q$ and $q \leq r$, then
$p \leq r$, (3) if $p \leq q$ and $q \leq p$, then $p=q$,
for all $p,q,r \in L$). Furthermore, $L$
contains distinguished elements $0$ and $1$ satisfying
$0 \leq p$ and $p \leq 1$, for all $p \in L$.
Finally, $L$ is endowed with a function $'$
({\em orthocomplementation}) from $L$ to $L$ satisfying the
conditions (1) $p''=p$, (2) if $p \leq q$, then $q' \leq p'$,
(3) the least upper bound of $p$ and $p'$ exists and is $1$,
for all $p,q \in L$.
Note that these conditions imply $0'=1$, $1'=0$, and that the greatest
lower bound
of $p$ and $p'$ exists and is $0$, for all $p \in L$.

For example,
an arbitrary sublattice of the lattice of all closed linear subspaces of a
Hilbert space is an orthoposet, if it contains the
subspace $\{0\}$  and the full Hilbert space and is closed under the
orthogonal complement operation.
Namely, the subspace $\{0\}$ is the $0$ in the orthoposet, the full Hilbert
space
is the $1$, the set-theoretic inclusion is the ordering $\leq$,
and the orthogonal complement operation is the orthocomplementation $'$.

In the rest of this section we always assume that $L$ is an
arbitrary orthoposet.
We shall construct a Boolean algebra $B$ and an injective
mapping $\varphi:L\rightarrow B$ which preserves the order
relation and the orthocomplementation.
The construction goes essentially along the same lines as the
construction of Zierler and Schlessinger \cite{ZirlSchl-65}
and Katrno\v{s}ka \cite{katrnoska-82} and is similar to the
proof of the Stone representation theorem for Boolean algebras, cf.\ Stone
\cite{stone}.
It is interesting to note that for a finite orthoposet
the constructed Boolean algebra will be finite as well.

We call a nonempty subset $K$ of $L$ an {\em ideal} if for all $p,q\in L$:
\begin{enumerate}
\item  if $p\in K$, then $p'\not \in K$,
\item if  $p\le q$ and $q\in K$, then $p\in K $.
\end{enumerate}
Clearly, if $K$ is an ideal, then  $0\in K$. An ideal $I$ is {\em maximal}
provided that if $K$ is an
ideal and $I\subseteq K$, then $K=I$.

Let ${\cal I}$ be the set of all maximal ideals in $L$, and let $B$ be the
power set of ${\cal I}$ considered as a Boolean algebra, i.e.\ $B$ is the
Boolean algebra which consists of all subsets of ${\cal I}$. The order
relation in $B$ is the set-theoretic inclusion, the ortholattice operations
{\em complement}, {\em or}, and {\em and} are given by the set-theoretic
complement,  union, and  intersection, and the elements $0$ and $1$ of the
Boolean algebra are just the empty set and the full set ${\cal I}$.
Consider the map
\[ \varphi:L \to B \]
which maps each element $p \in L$ to the set
\[   \varphi(p) = \{ I \in {\cal I} \ | \ p \not\in I \} \]
of all maximal ideals which do not contain $p$.
We claim that the map $\varphi$
\begin{enumerate}
\item[(i)] is injective,
\item[(ii)]  preserves the order relation,
\item[(iii)]  preserves  complementation.
\end{enumerate}
This provides an embedding of quantum logic into classical logic which
preserves the implication relation and the negation.\footnote{Note that for
a finite orthoposet $L$
the Boolean algebra $B$ is finite as well. Indeed, if $L$ is finite, then
it has only finitely many
subsets, especially only finitely many maximal ideals. Hence ${\cal I}$ is
finite, and thus also
its power set $B$ is finite.}

The rest of this section consists of the proof of the three claims above.
Let us start with claim (ii). Assume that $p,q \in L$ satisfy
$p \leq q$. We have to
show the inclusion
\[ \varphi(p) \subseteq \varphi(q) \,.\]
Take a maximal ideal $I \in \varphi(p)$. Then $p \not\in I$. If $q$ were
contained in $I$, then by condition 2. in
the definition of an ideal also $p$ had to be
contained in $I$. Hence $q \not\in I$, thus proving that $I \in \varphi(q)$.

Before we come to claims (iii) and (i) we give another
characterization of maximal ideals. We start with the following
assertion which will also be needed later:
\begin{equation} \label{auxideal}
  \begin{array}{l}
\mbox{\rm If $I$ is an ideal and $r \in L$ with $r \not\in I$
    and $r^\prime\not\in I$,} \\
\mbox{\rm then also the set $J = I \cup \{s \in L \mid s \leq r\}$ is an ideal.}
  \end{array}
\end{equation}
Here is the proof: It is clear that $J$ satisfies condition 2. in the
definition of an ideal. To show
that it satisfies condition 1. assume to the contrary that there exists
$s\in J$ and $s'\in J$, for some $s\in L$.
Then one of the following conditions  must be true;
(I) $s,s'\in J$,
(II) $s\le r$ and $s'\le r$,
(III) $s\in I$ and $s'\le r$,
(IV) $s\le r$, $s'\in I$.
The first case is impossible since $I$ is an ideal.
The second case is ruled out by the fact that $r\neq 1$
(namely, $r=1$ would imply $r'=0$ which would contradict our assumption $r'\not\in I$).
The third case is impossible since $s'\le r$ implies $r'\le s$ which,
combined with $s\in I$ would imply $r'\in I$, contrary to our assumption.
Finally the fourth case is nothing but a reformulation
of the third case with $s$ and $s'$ interchanged.
Thus we have proved that $J$ is an ideal and have proved the assertion
(\ref{auxideal}).

Next, we prove the following new characterization of maximal ideals:
\begin{equation} \label{maxideal}
\mbox{\rm An ideal $I$ is a maximal ideal iff $r \not\in I$ implies $r^\prime \in I$.}
\end{equation}
To prove this first assume that for all $r\in L$, if $r\not \in I$,
then $r'\in I$ and suppose  $I$ is a {\em proper} subset of an ideal $K$.
Then there exists $p\in K$ such that $p\not \in I$.
By our hypothesis
(for all $r\in L$, $r\not \in I$ implies $r'\in I$),
we have  $p^\prime\in I$. Thus both $p\in K$ and
$p^\prime\in K$. This contradicts the fact that $K$ is an ideal.

Conversely, suppose that $I$ is a maximal ideal in $L$ and suppose,
to the contrary, that for some
$r\in L,$
\begin{equation}\label{one}
r\not\in I\ {\rm and}\ r'\not\in I\,.
\end{equation}
Of course $r\neq 1$, since $1'=0\ {\rm and}\ 0\in I$. Let
\begin{equation}\label{two}
J=I\cup(r)
\end{equation}
where $(r)=\{s\in L \ |\ s\le r\}$ is the principal ideal of $r$
(note that $(r)$ is indeed an ideal). Then,
under assumption (\ref{one}), using (\ref{auxideal}) above, we have that
$J$ is an ideal which properly contains $I$. This contradicts
the maximality of $I$ and
ends the proof of the assertion (\ref{maxideal}).

For claim (iii) we have to show the relation:
\[  \varphi(p') = {\cal I} \setminus \varphi(p)\,, \]
for all $p \in L$. This can be restated as
\[ I \in \varphi(p') \; {\rm iff} \; I \not\in \varphi(p) \]
for all $I \in {\cal I}$.
But this means $p' \not\in I \; {\rm iff} \; p \in I$, which
follows directly from condition 1. in the definition of an ideal and
from assertion (\ref{maxideal}).

We proceed to claim (i), which states that $\varphi$ is injective,
i.e., if $p \neq q$, then $\varphi(p)\neq \varphi(q)$.
But $p\neq q$ is equivalent to $p\not\le q\
{\it or}\ q\not\le p$.
 Furthermore, if we can show that there is a maximal ideal $I$ such that
$q\in I$ and $p\not\in I$
then it follows easily that $\varphi(p)\neq\varphi(q)$.
Indeed, $p\not\in I$ means $I\in \varphi(p)$
and  $q\in I$ means $I\not\in\varphi(q)$.
It is therefore enough to prove that:
\begin{quote}
If $p\not\le q$, then there exists a maximal ideal $I$
such that $q\in I$ and $p\not\in I$.
\end{quote}
To prove this we note that since $p\not\le q$, we have $p\neq 0$. Let
\[ {\cal I}_{pq}=\{K\subseteq L \mid K\ {\rm is\ an\ ideal\ and}\ p\not\in
K\ {\rm and}\ q\in K\}.\]
We have to show that among the elements of ${\cal I}_{p,q}$ there is a
maximal ideal.
Therefore we will use Zorn's Lemma.
In order to apply it to ${\cal I}_{p,q}$ we have to show
 that ${\cal I}_{p,q}$ is not empty and
that every chain in ${\cal I}_{p,q}$ has an upper bound.

The set ${\cal I}_{p,q}$ is not empty since $(q) \in {\cal I}_{p,q}$.
Now we are going to show
that every chain in ${\cal I}_{p,q}$ has an upper bound.
This means that, given a subset ({\em chain}) ${\cal C}$ of
${\cal I}_{p,q}$ with the property
\[ {\rm for \mbox{ } all} \; J,K \in {\cal C} \;
{\rm one \mbox{ }  has} \; J \subseteq  K \ \mbox{\rm or } K \subseteq J \,, \]
we have to show that there is an element ({\em upper bound})
$U  \in {\cal I}_{p,q}$ with
$K \subseteq U$ for all $K \in {\cal C}$. The union
\[ U_{\cal C} = \bigcup_{K \in {\cal C}} K \]
of all ideals $K \in {\cal C}$ is the required upper bound!
It is clear that all $K \in {\cal C}$ are subsets of $U_{\cal C}$.
We have to show
that $U_{\cal C}$ is an element of ${\cal I}_{p,q}$. Since $p \not\in K$ for all
$K \in {\cal C}$ we also have $p \not\in U_{\cal C}$. Similarly, since
$q \in K$ for some (even all) $K \in {\cal C}$, we have $q \in U_{\cal C}$.
We still have to show that $U_{\cal C}$ is an ideal.
Given two propositions $r,s$ with $r \leq s$ and $s \in U_{\cal C}$ we
conclude that $s$ must be contained in one of the ideals $K \in {\cal C}$.
Hence also $r \in K \subseteq U_{\cal C}$. Now assume $r \in U_{\cal C}$.
Is it possible that  the complement $r'$ belongs to $U_{\cal C}$? The
answer is negative, since otherwise
$r \in J$ and $r' \in K$, for some ideals $J,K \in {\cal C}$. But since
${\cal C}$ is a chain we have $J \subseteq K$ or $K \subseteq J$, hence
$r, r' \in K$ in the first case and $r, r' \in J$ in the second case. Both
cases contradict the fact that $J$ and $K$ are ideals.
Hence, $U_{\cal C}$ is an ideal and thus an element of ${\cal I}_{p,q}$.
We have proved that ${\cal I}_{p,q}$ is not empty and that each chain in
${\cal I}_{p,q}$ has an upper bound in ${\cal I}_{p,q}$.

Consequently, we can apply Zorn's Lemma to ${\cal I}_{p,q}$ and obtain a maximal
element $I$ in the ordered set ${\cal I}_{p,q}$.
Thus
\begin{equation}\label{four}
p\not\in I\ {\rm and}\ q\in I.
\end{equation}
 It remains to show that $I$ is a maximal ideal in $L$. Thus suppose, to
the contrary, that $I$
is {\em not\/} a maximal ideal in $L$.

By (\ref{maxideal}) there exists $r\in L$ such that both
$r\not\in I$ and $r'\not\in I$.
Furthermore, since $p\neq 0$, then either $p\not\le r$ or $p\not\le r'$.
Without
loss of generality suppose
\begin{equation}\label{five}
p\not\le r.
\end{equation}
It follows, by (\ref{auxideal}), and since $r\not\in I$ and
$r'\not\in I$, that $I\cup(r)$ is an
ideal properly containing $I$.  But since, by Conditions (\ref{four}) and
(\ref{five}), $q\in I$ and
$p\not\le r$, we have
\begin{center}
$p\not\in I\cup(r)$ and $q\in I\cup(r)$.
\end{center}
Thus $I\cup(r)\in {\cal I}_{pq}$ and, since $r\not\in I$, we deduce
that $I\cup(r)$ properly contains $I$,
contradicting the fact that $I$ is a {\em maximal element} in ${\cal
I}_{pq}$.
This ends the proof of claim (i),
the claim that the map $\varphi$ is injective.

We have shown:
\begin{quote}
{\em Any orthoposet can be embedded into a Boolean algebra where the
embedding preserves the order relation and the complementation.}
\end{quote}

\subsection{Injective order preserving morphisms}
\label{section:injorder}


In this section we analyze a different embedding suggested by Malhas
\cite{malhas-87,malhas-92}.

We consider
an orthocomplemented lattice $(L, \leq, 0, 1, ')$, i.e.\
a lattice $(L, \leq, 0, 1)$ with $0\leq x\leq 1$
for all $x\in L$, with  orthocomplementation, that is with a mapping
$' : L \rightarrow L$ satisfying the following three properties: a) $x''=x$,
b) if $x\leq y$, then $y'\leq x'$, c) $x\cdot x'=0$ and $y \vee y'=1$. Here
$x\cdot y= {\rm glb} (x,y)$ and $x \vee y = {\rm lub}(x,y) $.

Furthermore, we will assume that $L$ is atomic\footnote{For every
$x\in L\setminus\{0\}$, there is an atom $a\in L$ such that $a\leq x$.
An atom is an element $a\in L$ with the property that if $0\leq y \leq a$,
then $y=0$ or $y=a$.} and satisfies the following additional property:

\begin{equation}
\label{*}
\mbox{ for all }\; x,y \in L, x \leq y \; \mbox{ iff for every atom }\;
a\in L, a\leq x\;
\mbox{ implies} \; a\leq y.
\end{equation}

Every atomic Boolean algebra and the lattice of closed subspaces of a separable
Hilbert space satisfy the above conditions.

Consider next a set $U$\footnote{Not containing the logical symbols $\cup,
', \rightarrow $.}
 and let $W(U)$ be the smallest set of words over the alphabet
$U \cup \{', \rightarrow\}$ which contains $U$ and
is closed under negation (if $A\in W(U)$, then $ A' \in W(U)$) and
implication (if
$A, B
\in W(U)$, then
$A \rightarrow B \in W(U)$).\footnote{Define in a natural way $A \cup B =
A' \rightarrow B$,
$A \cap B =  (A \rightarrow  B')'$, $ A \leftrightarrow B = (A \rightarrow
B) \cap
(B \rightarrow A)$.} The elements of $U$ are called {\it simple
propositions} and
the elements of $W(U)$ are called {\it (compound) propositions}.

A {\it valuation} is a mapping
\[t: W(U)  \rightarrow {\bf 2}\]
such that $t(A) \not= t(A')$ and $t(A \rightarrow B) = 0$ iff $t(A)=1$ and
$t(B)=0$.
Clearly, every assignment $s: U \rightarrow {\bf 2}$ can be extended to a
unique valuation $t_s$.

A {\it tautology} is a proposition $A$ which is true under every possible
valuation,
i.e., $t(A)=1$, for every valuation $t$. A set $ {\cal K} \subseteq W(U)$
is {\it consistent} if there is a valuation making true every proposition
in $\cal {K}$. Let $A\in W(U)$ and ${\cal K} \subseteq W(U)$.
We say that $A$ {\it derives} from $ {{\cal K}}$, and write
${\cal K}\models A$, in case  $t(A)=1$ for each valuation $t$ which makes
true every proposition in
${\cal K} $ (that is, $t(B)=1$, for all $B\in {\cal K}$).
We define the set of consequences of ${\cal K}$ by
\[Con({\cal K}) = \{A\in W(U) \mid {\cal K}\models A\}.\]
Finally, a set ${\cal K}$ is a {\it theory} if ${\cal K}$ is a fixed-point
of the operator
$Con$:
\[Con({\cal K}) = {\cal K}.\]
 It is easy to see that $Con$ is in fact a finitary closure operator, i.e.,
it satisfies the
following four properties:
\begin{itemize}
\item ${\cal K} \subseteq Con({\cal K})$,
\item if $ {\cal K} \subseteq \tilde{\cal K}$, then
         $Con({\cal K})\subseteq Con(\tilde{\cal K})$,
\item $Con(Con({\cal K})) = Con({\cal K})$,
\item $Con({\cal K}) = \bigcup_{\{X\subseteq {\cal K}, X \;\mbox{
finite}\}} Con(X)$.
\end{itemize}
The first three properties can be proved easily. A topological proof for
the fourth property can be found in Appendix B.

The main example of a theory can be obtained by taking a set $X$ of valuations
and constructing the set of all propositions true under all valuations in $X$:
\[Th(X) = \{ A\in W(U)\mid t(A)=1, \; \mbox{ for all} \; t\in X\}.\]

In fact, every theory is of the above form, that is, {\it for every theory
${\cal K}$ there exists a set of valuations $X$ (depending upon ${\cal K}$)
 such that ${\cal K} = Th(X).$ }  Indeed, take
\[X_{{\cal K}} = \{ t: W(U) \rightarrow {\bf 2} \mid t \; \mbox{ valuation
with} \;
t(A)=1,\; \mbox{ for all }\; A\in {\cal K}\},\]
and notice that

\begin{eqnarray*}
Th(X_{{\cal K}}) & = & \{B\in W(U)\mid t(B)=1,\; \mbox{ for all }\; t\in
X_{{\cal K}}\}\\
& = & \{ B\in W(U)\mid t(B)=1,\; \mbox{ for every valuation with}\; t(A)=1,\\
&    &  \mbox{ for all }\; A\in {\cal K}\}\\
& = & Con({\cal K}) = {\cal K}.
\end{eqnarray*}

In other words, {\it theories are those sets of propositions which are true
under
a certain set of valuations (interpretations).}

Let now ${\cal T}$ be a theory. Two elements $p,q\in U$ are ${\cal
T}$-equivalent,
written $p \equiv_{{\cal T}} q$, in case $p \leftrightarrow q\in {\cal T}$. The
relation $\equiv_{{\cal T}}$ is an equivalence relation. The equivalence
class of
$p$ is $[p]_{{\cal T}}=\{q\in U\mid p \equiv_{{\cal T}} q\}$ and the factor
set is denoted
by $U_{\equiv_{{\cal T}}}$; for brevity, we will sometimes write $[p]$
instead of $[p]_{{\cal T}}$.  The factor set comes with a
natural partial order:
\[[p] \leq [q] \; \mbox{ if}\; p \rightarrow q\in {\cal T}.\]
Note that in general, $(U_{\equiv_{{\cal T}}}, \leq)$ is not a
Boolean algebra.\footnote{For instance, in case ${\cal T} = Con(\{p\})$,
for some $p\in U$. If $U$ has at least three elements, then
$(U_{\equiv_{{\cal T}}}, \leq)$ does not have a minimum.}

In a similar way we can define the $\equiv_{{\cal T}}$-equivalence of two
propositions:
\[ A \equiv_{{\cal T}} B \; \mbox{ if }\; A \leftrightarrow B \in {\cal T}.\]
Denote by $[[A]]_{{\cal T}}$ (shortly, $[[A]]$) the equivalence class of
$A$ and note that for every
$p\in U$,
\[ [p]=[[p]] \cap U.\]
The resulting Boolean algebra $W(U)_{\equiv_{{\cal T}}}$ is the Lindenbaum
algebra of ${\cal T}$.

Fix now an atomic orthocomplemented lattice $(L, \leq, 0, 1, ')$ satisfying
(\ref{*}).
Let $U$ be a set of cardinality greater or equal to $L$ and fix a
surjective mapping
$f:  U \rightarrow L$. For every atom $a\in L$, let $s_a: U \rightarrow
{\bf 2}$ be
the assignment defined by $s_a (p)=1$ iff $a\leq f(p)$. Take
\[X = \{t_{s_a} \mid a \; \mbox{ is an atom of}\; L\}\footnote{Recall that
$t_s$ is the unique valuation extending $s$.}
\; \mbox{\rm and } {\cal T} = Th(X) . \]

 Malhas \cite{malhas-87,malhas-92} has proven that the {\it lattice
$(U_{\equiv_{{\cal T}}}, \leq)$ is orthocomplemented,} and, in fact,
{\it isomorphic to $L$}.
Here is the argument. Note first that there exist two elements
$\underline{0}, \underline{1}$ in $U$ such that
$f(\underline{0})=0, \; f(\underline{1})=1$. Clearly, $\underline{0}\not\in
{\cal T}$,
but $\underline{1}\in {\cal T}$. Indeed, for every atom $a$, $a\leq
f(\underline{1})=1$, so
$s_a(\underline{1}) = 1$, a.s.o.

Secondly, for every $p,q\in U$,
\[p \rightarrow q \in {\cal T} \; \mbox{ iff } \; f(p) \leq f(q).\]

If $p \rightarrow q\not\in {\cal T}$, then there exists an atom $a\in L$
such that
$t_{s_a}(p \rightarrow q)=0,$ so $s_a(p) = t_{s_a}(p)=1, \;s_a(q) =
t_{s_a}(q)=0$, which---according to the definition of $s_a$---mean
$a\leq f(p)$, but $a\not\leq f(q)$. If $f(p) \leq f(q)$, then
$a\leq f(q)$, a contradiction.  Conversely, if $f(p) \not\leq f(q)$, then
by (\ref{*}) there
exists an atom $a$ such that $a\leq f(p)$ and $a\not\leq f(q)$. So, $s_a(p)
= t_{s_a}(p)=1,
\;s_a(q) = t_{s_a}(q)=0$, i.e., $(p\rightarrow q)\not\in {\cal T}$.

As immediate consequences we deduce the validity of the following three
relations: for all
$p,q\in U$,
\begin{itemize}
\item  $ f(p) \leq f(q)$ iff  $[p]\leq [q]$,
\item $ f(p) = f(q)$ iff  $[p] = [q]$,
\item  $[\underline{0}] \leq [p] \leq [\underline{1}]$.
\end{itemize}

Two simple propositions $p,q\in U$ are {\it conjugate} in case
$f(p)'=f(q)$.\footnote{Of course,
this relation is symmetrical.} Define now the operation $^{*}: U_{{\cal T}}
\rightarrow U_{{\cal
T}}$ as follows: $[p]^{*}=[q]$ in case $q$ is a conjugate of $p$.  It is
not difficult to see that
the operation $^{*}$ is well-defined and actually is an
orthocomplementation. It follows that
$(U_{{\cal T}}, \leq_{{\cal T}}, ^{*})$ is an orthocomplemented lattice.

To finish the argument we will show that this lattice is {\it isomorphic}
with $L$. The
isomorphism is given by the mapping $\psi : U_{{\cal T}} \rightarrow L$
defined by
the formula $\psi([p]) =  f(p)$. This is a well-defined function
(because $ f(p) = f(q)$ iff  $[p]=[q]$), which is bijective ($\psi([p])=
\psi([q])$
implies $ f(p) = f(q)$, and surjective because $f$ is onto). If $[p]\leq [q]$,
then $f(p)\leq f(q)$, i.e. $\psi([p]) \leq \psi([q])$. Finally, if
$q$ is a conjugate
of $p$, then
\[\psi([p]^{*}) = \psi([q]) = f(q) = f(p)'= \psi([p])'.\]

In particular, {\it there exists a theory
whose induced
orthoposet is isomorphic to the lattice of all closed subspaces of
a separable Hilbert
space}. How does this relate to the Kochen-Specker theorem?
{\it The natural embedding
\[ \Gamma: U_{\equiv_{{\cal T}}} \rightarrow W(U)_{\equiv_{{\cal T}}}, \
\Gamma ([p]) = [[p]]\]
is order preserving and one-to-one, but in general it does not preserve
orthocomplementation}, i.e.\ in general
$\Gamma ([p]^{*}) \neq \Gamma ([p])'$. We always have $\Gamma ([p]^{*}) \leq
\Gamma ([p])'$, but sometimes $
\Gamma ([p])'\not\leq \Gamma ([p]^{*})$.
The reason is that for every pair of conjugate simple propositions $p,q$
one has $(p \rightarrow q')\in {\cal T}$, but the converse is not true.

By combining the inverse $\psi^{-1}$ of the isomorphism $\psi$ with $\Gamma$
we obtain an embedding $\varphi$ of $L$ into the Boolean Lindenbaum algebra
$W(U)_{\equiv_{{\cal T}}}$. Thus,
the above construction of Malhas gives us another method how {\em to embed
any quantum logic
into a Boolean logic in case we require that only the  order is
preserved}.\footnote{In Section \ref{section:2.3} we saw that it is possible to
embed quantum
logic into a Boolean logic preserving the order and the complement.}

Next we shall give a simple example of a Malhas type embedding
$\varphi : MO_2 \rightarrow {\bf 2}^4$. Consider
again the finite
quantum logic $MO_2$ represented in  Figure \ref{f-hd-mo2}.
Let us choose
$$U=\{A,B,C,D,E,F,G,H\}.$$
Since $U$ contains more elements than $MO_2$, we can map $U$ surjectively
onto $MO_2$; e.g.,
\begin{eqnarray*}
f(A)&=& 0,\\
f(B)&=& p_-,\\
f(C)&=& p_-,\\
f(D)&=& p_+,\\
f(E)&=& q_-,\\
f(F)&=& q_+,\\
f(G)&=& 1,\\
f(H)&=& 1.
\end{eqnarray*}

For every atom  $a \in MO_2$, let us introduce the truth assignment
$s_{a}:U\rightarrow {\bf 2}=\{0,1\}$ as defined above
(i.e.\ $s_a(r)=1$ iff $a \leq f(r)$) and
thus a valuation on $W(U)$ separating it from the rest of the atoms of
$MO_2$. That is, for instance, associate with $p_-\in MO_2$ the function
$s_{p_-}$
 as follows:
\begin{eqnarray*}
&&s_{p_-}(A)=s_{p_-}(D)=s_{p_-}(E)=s_{p_-}(F)=0,\\
&&s_{p_-}(B)=s_{p_-}(C)=s_{p_-}(G)=s_{p_-}(H)=1.
\end{eqnarray*}
The truth assignments associated with all the atoms are listed
in Table \ref{t-mo2-tvs2}.
\begin{table}
\begin{center}
\begin{tabular}{|c|cccccccc|}
\hline\hline
      &$A$& $B$& $C$& $D$& $E$& $F$& $G$& $H$\\
\hline
$s_{p_{-}}$ &0  &  1&     1 &   0 &0  &  0&     1 &   1 \\
$s_{p_{+}}$ &0  &  0&     0 &   1 &0  &  0&     1 &   1 \\
$s_{q_{-}}$ &0  &  0&     0 &   0 &1  &  0&     1 &   1 \\
$s_{q_{+}}$ &0  &  0&     0 &   0 &0  &  1&     1 &   1 \\
\hline\hline
\end{tabular}
\end{center}
\caption{Truth assignments on $U$ corresponding to
atoms $p_{-},p_{+},q_{-},q_{+}\in MO_{2}$.
\label{t-mo2-tvs2}}
\end{table}

\begin{figure}
\begin{center}
\unitlength 0.85mm
\linethickness{0.4pt}
\begin{picture}(113.00,83.00)
\put(60.00,80.00){\circle*{2.11}}
\put(30.00,60.00){\circle*{2.11}}
\put(50.00,60.00){\circle*{2.11}}
\put(70.00,60.00){\circle*{2.11}}
\put(90.00,60.00){\circle*{2.11}}
\put(30.00,40.00){\circle*{2.11}}
\put(30.00,20.00){\circle*{2.11}}
\put(50.00,40.00){\circle*{2.11}}
\put(50.00,20.00){\circle*{2.11}}
\put(70.00,40.00){\circle*{2.11}}
\put(70.00,20.00){\circle*{2.11}}
\put(90.00,40.00){\circle*{2.11}}
\put(90.00,20.00){\circle*{2.11}}
\put(60.00,0.00){\circle*{2.11}}
\put(10.00,40.00){\circle*{2.11}}
\put(110.00,40.00){\circle*{2.11}}
\put(60.00,0.00){\line(-3,2){30.00}}
\put(30.00,20.00){\line(-1,1){20.00}}
\put(10.00,40.00){\line(1,1){20.00}}
\put(30.00,60.00){\line(3,2){30.00}}
\put(60.00,80.00){\line(3,-2){30.00}}
\put(90.00,60.00){\line(1,-1){20.00}}
\put(110.00,40.00){\line(-1,-1){20.00}}
\put(90.00,20.00){\line(-3,-2){30.00}}
\put(60.00,0.00){\line(-1,2){10.00}}
\put(50.00,20.00){\line(-2,1){40.00}}
\put(10.00,40.00){\line(2,1){40.00}}
\put(50.00,60.00){\line(2,-1){40.00}}
\put(90.00,40.00){\line(-2,-1){40.00}}
\put(50.00,20.00){\line(1,1){40.00}}
\put(90.00,60.00){\line(0,-1){40.00}}
\put(90.00,20.00){\line(-2,1){40.00}}
\put(50.00,40.00){\line(0,1){20.00}}
\put(50.00,60.00){\line(1,2){10.00}}
\put(60.00,80.00){\line(1,-2){10.00}}
\put(70.00,60.00){\line(-1,-1){40.00}}
\put(30.00,20.00){\line(0,1){40.00}}
\put(30.00,60.00){\line(2,-1){40.00}}
\put(60.00,0.00){\line(1,2){10.00}}
\put(70.00,20.00){\line(-2,1){40.00}}
\put(30.00,40.00){\line(2,1){40.00}}
\put(70.00,60.00){\line(2,-1){40.00}}
\put(110.00,40.00){\line(-2,-1){40.00}}
\put(70.00,20.00){\line(0,1){20.00}}
\put(70.00,0.00){\makebox(0,0)[lc]{$0\equiv [[A]]$}}
\put(70.00,80.00){\makebox(0,0)[lc]{$1\equiv [[G]]=[[B\vee D\vee E\vee F]]$}}
\put(30.00,13.00){\makebox(0,0)[cc]{$[[B]]$}}
\put(50.00,13.00){\makebox(0,0)[cc]{$[[D]]$}}
\put(70.00,13.00){\makebox(0,0)[cc]{$[[E]]$}}
\put(90.00,13.00){\makebox(0,0)[cc]{$[[F]]$}}
\put(10.00,35.00){\makebox(0,0)[cc]{$[[B\vee D]]$}}
\put(31.33,35.00){\makebox(0,0)[cc]{$[[B\vee E]]$}}
\put(50.00,35.00){\makebox(0,0)[cc]{$[[B\vee F]]$}}
\put(71.00,35.00){\makebox(0,0)[cc]{$[[D\vee E]]$}}
\put(91.00,35.00){\makebox(0,0)[cc]{$[[D\vee F]]$}}
\put(110.00,35.00){\makebox(0,0)[cc]{$[[E\vee F]]$}}
\put(90.00,65.00){\makebox(0,0)[lc]{$[[D\vee E\vee F]]$}}
\put(75.00,65.00){\makebox(0,0)[cc]{$[[B\vee E\vee F]]$}}
\put(45.00,65.00){\makebox(0,0)[cc]{$[[B\vee D\vee F]]$}}
\put(30.00,65.00){\makebox(0,0)[rc]{$[[B\vee D\vee E]]$}}
\put(60.00,80.00){\circle{6.00}}
\put(60.00,0.00){\circle{6.00}}
\put(30.00,20.00){\circle{6.00}}
\put(50.00,20.00){\circle{6.00}}
\put(70.00,20.00){\circle{6.00}}
\put(90.00,20.00){\circle{6.00}}
\end{picture}
\end{center}
\caption{\label{f-pt-tre}Hasse diagram of an embedding of the
quantum logic $MO_2$ represented by Figure \protect\ref{f-hd-mo2}.
Concentric circles indicate the embedding.}
\end{figure}
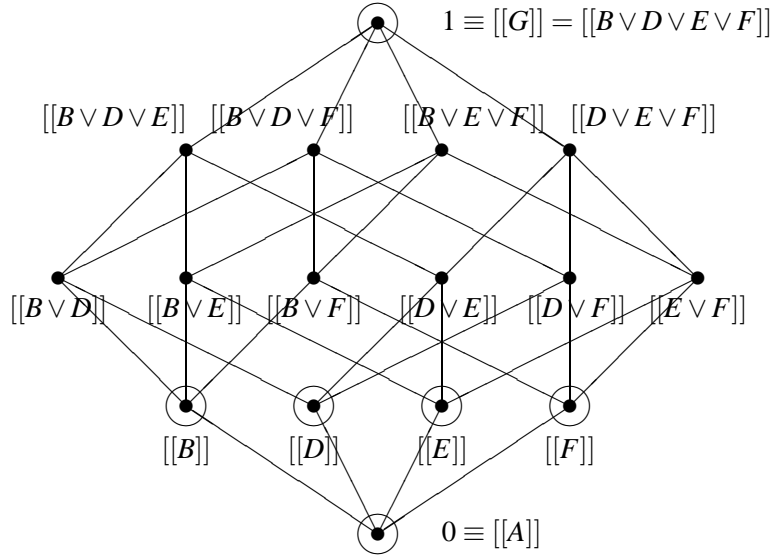

The theory ${\cal T}$ we are thus dealing with is determined by the union
of all the truth
assignments; i.e.,
\[ X =\{t_{s_{p_-}},t_{s_{p_+}},t_{s_{q_-}},t_{s_{q_+}}\} \ \mbox{\rm and }
     {\cal T} = Th(X) .\]
The way it was constructed, $U$ splits into six equivalence classes with
respect to the theory ${\cal T}$; i.e.,
\[ U_{\equiv_{{\cal T}}}=\{[A],[B],[D],[E],[F],[G]\}. \]
Since $[p]\rightarrow [q]$ if and only if $(p\rightarrow q)\in {\cal T}$,
we obtain
a partial order on $U_{\equiv_{{\cal T}}}$ induced by $T$ which
isomorphically reflects the
original quantum logic $MO_2$.
The Boolean Lindenbaum algebra $W(U)_{\equiv_{{\cal T}}}={\bf 2}^4$ is
obtained by forming all the compound
propositions of $U$ and imposing a partial order with respect to ${\cal
T}$. It  is represented in Figure \ref{f-pt-tre}.
The embedding is  given by
\begin{eqnarray*}
\varphi (0)&=&[[A]],\\
\varphi( p_-)&=&[[B]],\\
\varphi( p_+)&=&[[D]],\\
\varphi(q_-)&=&[[E]],\\
\varphi (q_+)&=&[[F]],\\
\varphi (1)&=&[[G]] .
\end{eqnarray*}
It is order--preserving but does not preserve operations
such as the complement. Although, in this particular example,
$f(B)=(f(D))'$ implies
$(B\rightarrow  D')\in {\cal T}$, the converse is not true in general.
For example, there is no $s\in X$ for which $s(B)=s(E)=1$.
Thus, $(B\rightarrow E')\in T$, but $f(B)\neq (f(E))'$.

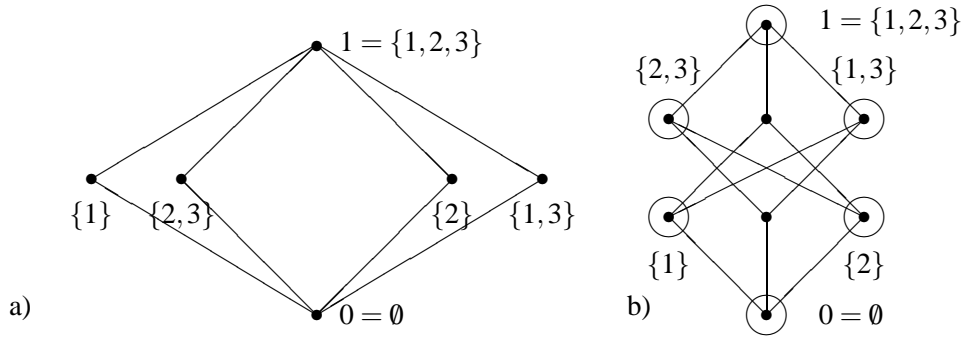
\begin{figure}\begin{center}
a)
\unitlength 0.60mm
\linethickness{0.4pt}
\begin{picture}(111.06,60.73)
\put(60.00,0.00){\circle*{2.11}}
\put(30.00,30.00){\circle*{2.11}}
\put(60.00,59.67){\circle*{2.11}}
\put(90.00,30.00){\circle*{2.11}}
\put(60.00,0.00){\line(-1,1){30.00}}
\put(30.00,30.00){\line(1,1){30.00}}
\put(60.00,60.00){\line(1,-1){30.00}}
\put(90.00,30.00){\line(-1,-1){30.00}}
\put(65.00,0.00){\makebox(0,0)[lc]{$0=\emptyset$}}
\put(65.00,60.00){\makebox(0,0)[lc]{$1=\{1,2,3\}$}}
\put(30.00,22.00){\makebox(0,0)[cc]{$\{2,3\}$}}
\put(90.00,22.00){\makebox(0,0)[cc]{$\{2\}$}}
\put(60.00,0.00){\line(-5,3){50.00}}
\put(10.00,30.00){\line(5,3){50.00}}
\put(60.00,60.00){\line(5,-3){50.00}}
\put(110.00,30.00){\line(-5,-3){50.00}}
\put(10.00,30.00){\circle*{2.11}}
\put(10.00,22.00){\makebox(0,0)[cc]{$\{1\}$}}
\put(110.00,30.00){\circle*{2.11}}
\put(110.00,22.00){\makebox(0,0)[cc]{$\{1,3\}$}}
\end{picture}
$\quad$
$\quad$
b)
\unitlength 1.30mm
\linethickness{0.4pt}
\begin{picture}(21.33,30.17)
\put(0.33,10.00){\circle*{1.00}}
\put(10.33,10.00){\circle*{1.00}}
\put(20.33,10.00){\circle*{1.00}}
\put(0.33,20.00){\circle*{1.00}}
\put(10.33,20.00){\circle*{1.00}}
\put(20.33,20.00){\circle*{1.00}}
\put(10.33,29.67){\circle*{1.00}}
\put(10.33,0.00){\circle*{1.00}}
\put(10.33,0.00){\line(-1,1){10.00}}
\put(0.33,10.00){\line(1,1){10.00}}
\put(10.33,20.00){\line(0,1){10.00}}
\put(10.33,30.00){\line(1,-1){10.00}}
\put(20.33,20.00){\line(-2,-1){20.00}}
\put(10.33,0.00){\line(0,1){10.00}}
\put(10.33,10.00){\line(-1,1){10.00}}
\put(0.33,20.00){\line(1,1){10.00}}
\put(10.33,10.00){\line(1,1){10.33}}
\put(10.33,20.00){\line(1,-1){10.00}}
\put(20.33,10.00){\line(-2,1){20.00}}
\put(20.33,10.00){\line(-1,-1){10.00}}
\put(0.33,5.00){\makebox(0,0)[cc]{$\{1\}$}}
\put(20.33,5.00){\makebox(0,0)[cc]{$\{2\}$}}
\put(15.66,0.00){\makebox(0,0)[lc]{$0=\emptyset$}}
\put(0.33,25.00){\makebox(0,0)[cc]{$\{2,3\}$}}
\put(20.33,25.00){\makebox(0,0)[cc]{$\{1,3\}$}}
\put(15.66,30.00){\makebox(0,0)[lc]{$1=\{1,2,3\}$}}
\put(0.33,10.00){\circle{4.00}}
\put(20.33,10.00){\circle{4.00}}
\put(0.33,20.00){\circle{4.00}}
\put(20.33,20.00){\circle{4.00}}
\put(10.33,0.00){\circle{4.00}}
\put(10.33,29.67){\circle{4.00}}
\end{picture}
\end{center}
\caption{\label{mooreh-e2} Hasse diagram of an embedding of $MO_2$
drawn in a)
into ${\bf 2}^3$ drawn in b).
Again, concentric circles indicate points of ${\bf 2}^3$ included in
$MO_2$.}
\end{figure}

One needs not be afraid of order-preserving embeddings which are no lattice
morphisms, after all.
Even automaton logics (see Svozil \cite[Chapter 11]{svozil-93},
Schaller and Svozil \cite{schaller-92,schaller-95,schaller-96},
and Dvure{\v{c}}enskij, Pulmannov{\'{a}} and Svozil \cite{dvur-pul-svo})
can be embedded in this way.
Take again the lattice $MO_2$ depicted in Figure
\ref{f-hd-mo2}. A partition (automaton) logic realization is, for
instance,
$$\{\{\{1\},\{2,3\}\},\{\{2\},\{1,3\}\}\},$$
with
\begin{eqnarray*}
\{1\}&\equiv &p_-,\\
\{2,3\}&\equiv& p_+,\\
\{2\}&\equiv & q_-,\\
\{1,3\}&\equiv &q_+
,\end{eqnarray*}
respectively.
If we take $\{1\}$,$\{2\}$ and $\{3\}$ as atoms, then the Boolean algebra
${\bf 2}^3$
generated by all subsets of $\{1,2,3\}$ with the set theoretic inclusion
as order relation suggests itself as a candidate for an embedding. The
embedding  is quite trivially given by
$$\varphi (p)=p\in {\bf 2}^{3}.$$
The particular example considered above is
represented in Figure \ref{mooreh-e2}.
It is not difficult to check that the embedding satisfies the requirements
(i) and (ii), that is, it is injective and order preserving.

It is important to realize at that point that, although
different automaton partition logical structures may be isomorphic from a
logical point of
view (one-to-one translatable elements, order relations and operations),
they may
be very different with respect to their embeddability. Indeed, any two
distinct partition logics correspond to two distinct embeddings.

\begin{figure}
\begin{center}
\unitlength 0.85mm
\linethickness{0.4pt}
\begin{picture}(113.00,83.00)
\put(60.00,80.00){\circle*{2.11}}
\put(30.00,60.00){\circle*{2.11}}
\put(50.00,60.00){\circle*{2.11}}
\put(70.00,60.00){\circle*{2.11}}
\put(90.00,60.00){\circle*{2.11}}
\put(30.00,40.00){\circle*{2.11}}
\put(30.00,20.00){\circle*{2.11}}
\put(50.00,40.00){\circle*{2.11}}
\put(50.00,20.00){\circle*{2.11}}
\put(70.00,40.00){\circle*{2.11}}
\put(70.00,20.00){\circle*{2.11}}
\put(90.00,40.00){\circle*{2.11}}
\put(90.00,20.00){\circle*{2.11}}
\put(60.00,0.00){\circle*{2.11}}
\put(10.00,40.00){\circle*{2.11}}
\put(110.00,40.00){\circle*{2.11}}
\put(60.00,0.00){\line(-3,2){30.00}}
\put(30.00,20.00){\line(-1,1){20.00}}
\put(10.00,40.00){\line(1,1){20.00}}
\put(30.00,60.00){\line(3,2){30.00}}
\put(60.00,80.00){\line(3,-2){30.00}}
\put(90.00,60.00){\line(1,-1){20.00}}
\put(110.00,40.00){\line(-1,-1){20.00}}
\put(90.00,20.00){\line(-3,-2){30.00}}
\put(60.00,0.00){\line(-1,2){10.00}}
\put(50.00,20.00){\line(-2,1){40.00}}
\put(10.00,40.00){\line(2,1){40.00}}
\put(50.00,60.00){\line(2,-1){40.00}}
\put(90.00,40.00){\line(-2,-1){40.00}}
\put(50.00,20.00){\line(1,1){40.00}}
\put(90.00,60.00){\line(0,-1){40.00}}
\put(90.00,20.00){\line(-2,1){40.00}}
\put(50.00,40.00){\line(0,1){20.00}}
\put(50.00,60.00){\line(1,2){10.00}}
\put(60.00,80.00){\line(1,-2){10.00}}
\put(70.00,60.00){\line(-1,-1){40.00}}
\put(30.00,20.00){\line(0,1){40.00}}
\put(30.00,60.00){\line(2,-1){40.00}}
\put(60.00,0.00){\line(1,2){10.00}}
\put(70.00,20.00){\line(-2,1){40.00}}
\put(30.00,40.00){\line(2,1){40.00}}
\put(70.00,60.00){\line(2,-1){40.00}}
\put(110.00,40.00){\line(-2,-1){40.00}}
\put(70.00,20.00){\line(0,1){20.00}}
\put(70.00,0.00){\makebox(0,0)[cc]{$0$}}
\put(70.00,80.00){\makebox(0,0)[lc]{$1$}}
\put(30.00,15.00){\makebox(0,0)[cc]{$\{1\}$}}
\put(50.00,15.00){\makebox(0,0)[cc]{$\{2\}$}}
\put(70.00,15.00){\makebox(0,0)[cc]{$\{3\}$}}
\put(90.00,15.00){\makebox(0,0)[cc]{$\{4\}$}}
\put(10.00,35.00){\makebox(0,0)[cc]{$\{1,2\}$}}
\put(31.33,35.00){\makebox(0,0)[lc]{$\{1,3\}$}}
\put(50.00,35.00){\makebox(0,0)[cc]{$\{1,4\}$}}
\put(71.00,35.00){\makebox(0,0)[lc]{$\{2,3\}$}}
\put(91.00,35.00){\makebox(0,0)[lc]{$\{2,4\}$}}
\put(110.00,35.00){\makebox(0,0)[cc]{$\{3,4\}$}}
\put(90.00,65.00){\makebox(0,0)[lc]{$\{2,3,4\}$}}
\put(70.00,65.00){\makebox(0,0)[cc]{$\{1,3,4\}$}}
\put(50.00,65.00){\makebox(0,0)[cc]{$\{1,2,4\}$}}
\put(30.00,65.00){\makebox(0,0)[rc]{$\{1,2,3\}$}}
\put(60.00,80.00){\circle{6.00}}
\put(60.00,0.00){\circle{6.00}}
\put(10.00,40.00){\circle{6.00}}
\put(30.00,40.00){\circle{6.00}}
\put(90.00,40.00){\circle{6.00}}
\put(110.00,40.00){\circle{6.00}}
\end{picture}
\end{center}
\caption{\label{f-pt-treb}
Hasse diagram of an embedding of the partition logic
$\{\{\{1,2\},$  $\{3,4\}\},$  $ \{\{1,3\}, $  $\{2,4\}\}\}$
into ${\bf 2}^4$ preserving ortholattice operations among co--measurable
propositions.
Concentric circles indicate the embedding.}
\end{figure}
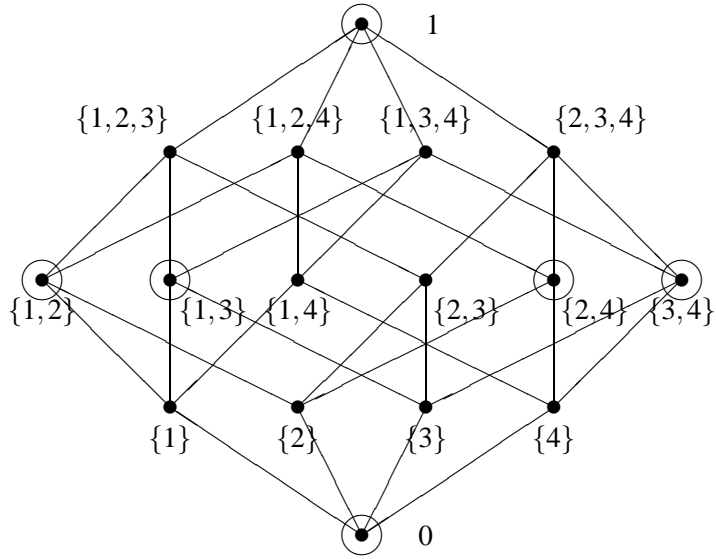

It should also be pointed out that in the case of an automaton partition
logic and for all finite subalgebras of the Hilbert lattice of
two-dimensional Hilbert space, it is always possible to find an
embedding corresponding to a logically equivalent partition
logic which is a lattice
morphism for co--measurable elements (modified requirement (iii)).
This is due to the fact that partition logics and $MO_{n}$ have a
separating set of valuations. In the $MO_2$ case, this is, for instance
$$\{\{\{1,2\},\{3,4\}\},\{\{1,3\},\{2,4\}\}\},$$
with
\begin{eqnarray*}
\{1,2\}&\equiv& p_-,\\
\{3,4\}&\equiv& p_+,\\
\{1,3\}&\equiv& q_-,\\
\{2,4\}&\equiv& q_+,
\end{eqnarray*}
respectively. This embedding is based upon the set of all
valuations listed in Table \ref{t-mo2-tvs}. These are exactly the mappings
from $MO_2$ to ${\bf 2}$ preserving the order relation and the complementation.
They correspond to the maximal ideals considered in Section 2.3. In this
special case the embedding is just the embedding obtained by applying
the construction of Section 2.3, which had been  suggested by
Zierler and Schlessinger \cite[Theorem 2.1]{ZirlSchl-65}.
\begin{table}
\begin{center}
\begin{tabular}{|c|cccc|}
\hline\hline
      &$p_{-}$& $p_{+}$& $q_{-}$& $q_{+}$\\
\hline
$s_1$ &1  &  0&     1 &   0  \\
$s_2$ &1  &  0&     0 &   1  \\
$s_3$ &0  &  1&     1 &   0  \\
$s_4$ &0  &  1&     0 &   1  \\
\hline\hline
\end{tabular}
\end{center}
\caption{The four valuations $s_{1},s_{2},s_{3},s_{4}$
on
$MO_2$ take on the values listed in the rows.
\label{t-mo2-tvs}}
\end{table}
The embedding is drawn in Figure \ref{f-pt-treb}.

\section{Surjective extensions?}

The original proposal put forward by EPR \cite{epr} in the last
paragraph of their paper was some form of completion  of quantum
mechanics. Clearly,  the first type of candidate for such a completion
is the sort of embedding reviewed above. The physical intuition behind
an embedding is that the ``actual physics'' is a classical one, but
because of some yet unknown reason, some of this ``hidden arena''
becomes observable while others remain hidden.

Nevertheless, there exists at least one other alternative to complete
quantum mechanics. This is best described by a {\em surjective map}
$\phi :B\rightarrow L$ of a classical Boolean algebra onto a quantum
logic, such that $\vert B\vert \ge \vert L\vert$.

Plato's cave metaphor applies to both approaches, in that observations
are mere shadows of some more fundamental entities.

\section{Summary}
We have reviewed several options for a classical ``understanding'' of
quantum mechanics. Particular emphasis has been given to
techniques for embedding quantum  universes into classical ones.
The term ``embedding'' is formalized here as usual. That is, an
embedding is a mapping of the entire set of quantum
observables into a (bigger) set of classical observables such that
different quantum observables correspond to different classical ones
(injectivity).

The term ``observables'' here is used for quantum propositions, some of
which (the complementary ones) might not be co--measurable,
see Gudder \cite{gudder1}.
It might
therefore be more appropriate to conceive these ``observables'' as
``potential observables.'' After a particular measurement has been
chosen, some of these observables are actually determined and others
(the complementary ones) become
``counterfactuals''  by quantum mechanical means; cf.\
Schr\"odinger's catalogue of expectation values
\cite[p.\ 823]{schrodinger}. For classical observables, there
is no distinction between ``observables'' and ``counterfactuals,''
because everything can be measured precisely, at least in principle.

We should mention also a {\it caveat}. The relationship between the
states of a quantum universe and the states of a classical universe into
which the former one is embedded is beyond the scope of this paper.

As might have been suspected, it turns out that, in order
to be able
to perform the mapping from the quantum universe into the classical one
consistently, important
structural elements of the quantum universe have to be sacrificed:
%
\begin{description}
\item[$\bullet$]
Since {\it per definition}, the quantum propositional calculus is
nondistributive (nonboolean), a straightforward embedding which
preserves all the logical operations among observables, irrespective of
whether or not they are co--measurable, is impossible.
This is due to the quantum mechanical feature of {\em complementarity}.
\item[$\bullet$]
One may restrict the preservation of the logical operations to be valid
only among mutually orthogonal propositions.
In this case it turns out that again
a consistent embedding is impossible, since no consistent meaning can be
given to the classical existence of ``counterfactuals.''
This is due to the quantum mechanical feature of {\em contextuality}.
That is,  quantum observables may appear different, depending on the way
by which they were measured (and inferred).
\item[$\bullet$]
In a further step, one may abandon preservation of lattice operations
such as {\it not} and the binary
{\it and} and
{\it or} operations altogether. One may merely require the preservation
of the
implicational structure (order relation). It
turns out that, with these provisos, it is indeed possible to map
quantum universes into classical ones. Stated differently, definite
values can be associated with elements of physical reality, irrespective
of whether they have been measured or not.
In this sense, that is, in terms of more
``comprehensive'' classical universes (the hidden parameter models),
 quantum mechanics can be ``understood.''
\end{description}

At the moment we can neither say if the nonpreservation of the binary
lattice operations (interpreted as {\it and} and {\it or}) is a
too high price for value definiteness, nor can we speculate of whether
or not the entire program of embedding quantum universes into classical
theories is a progressive or a degenerative case (compare Lakatosch
\cite{lakatosch}).

\appendix
\section*{Appendix A: Proof of the geometric lemma}

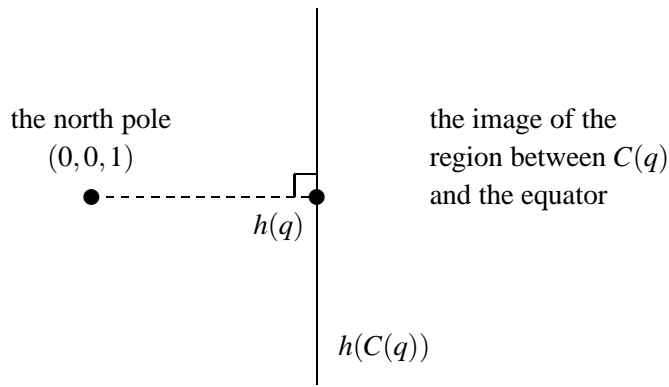
\begin{figure}[htbp]
\setlength{\unitlength}{1cm}
\begin{center}
\begin{picture}(8,5)
   \thinlines
   \put(2,2.5){\circle*{0.2}}
   \put(5,2.5){\circle*{0.2}}
   \put(4.7,2.8){\line(1,0){0.3}}
   \put(4.7,2.5){\line(0,1){0.3}}
   \multiput(2,2.5)(0.2,0){15}{\line(1,0){0.1}}
   \put(5,0){\line(0,1){5}}
   \put(2,3.5){\makebox(0,0){the north pole}}
   \put(2,3){\makebox(0,0){$(0,0,1)$}}
   \put(6.5,3.5){\makebox(0,0)[l]{the image of the}}
   \put(6.5,3){\makebox(0,0)[l]{region between $C(q)$}}
   \put(6.5,2.5){\makebox(0,0)[l]{and the equator}}
   \put(4.5,2.1){\makebox(0,0){$h(q)$}}
   \put(5.9,0.5){\makebox(0,0){$h(C(q))$}}
\end{picture}
\end{center}
\caption{The plane $H$ viewed from above.}
\label{figure:projectionh}
\end{figure}

In this appendix we are going to prove the geometric lemma
due to Piron \cite{piron-76} which was formulated in Section 2.2.
First let us restate it. Consider a point $q$ in the northern hemisphere
of the unit sphere $S^2 = \{p \in {\Bbb R}^3 \ | \ ||p||=1\}$. By
$C(q)$ we denote the unique
great circle which contains $q$ and the points
$\pm(q_y,-q_x,0)/\sqrt{q_x^2+q_y^2}$ in the equator, which are orthogonal
to $q$, compare Figure \ref{figure:greatcircle}.
We say that a point $p$ in the northern hemisphere {\em can be reached}
from a point $q$ in the
northern hemisphere, if there is a finite
sequence of points $q=q_0, q_1, \ldots, q_{n-1}, q_n=p$
in the northern hemisphere such that $q_i\in C(q_{i-1})$
for $i=1,\ldots,n$. The lemma states:
\begin{quote}
{\it If $q$ and $p$ are points in the northern hemisphere
with $p_z < q_z$, then $p$ can be reached from $q$.}
\end{quote}
For the proof we follow
Cooke, Keane, and Moran \cite{c-k-m} and Kalmbach \cite{kalmbach-86}).
We consider the
tangent plane $H=\{p \in {\Bbb R}^3\ | \ p_z=1\}$ of the unit sphere
in the north pole  and the projection $h$ from the northern hemisphere
onto this plane which maps each point
$q$ in the northern hemisphere to the intersection $h(q)$
of the line
through the origin and $q$ with the plane $H$.
This map $h$ is a bijection.
The north pole $(0,0,1)$ is mapped
to itself. For each $q$
in the northern
hemisphere (not equal to the north pole)
the image $h(C(q))$ of the
great
circle $C(q)$ is the line in $H$ which goes through $h(q)$
and is
orthogonal to the line through the north pole  and
through $h(q)$.
Note
that $C(q)$ is the intersection of a plane with $S^2$, and
$h(C(q))$ is the
intersection of the same plane with $H$;
see Figure
\ref{figure:projectionh}.
The line $h(C(q))$
divides
$H$ into two half planes. The half plane not containing the north
pole
is the image of the region in the northern
hemisphere between $C(q)$
and the equator.
Furthermore note that $q_z > p_z$ for two points in
the
northern hemisphere if and only if $h(p)$ is further
away from the north
pole than $h(q)$.
We proceed in two steps.

\begin{figure}[htbp]
\setlength{\unitlength}{1cm}
\begin{center}
\begin{picture}(9,6)
   \thinlines
   \put(2,2.5){\circle*{0.2}}
   \put(5,2.5){\circle*{0.2}}
   \put(4.7,2.8){\line(1,0){0.3}}
   \put(4.7,2.5){\line(0,1){0.3}}
   \multiput(2,2.5)(0.2,0){15}{\line(1,0){0.1}}
   \put(5,0){\line(0,1){6}}
   \put(1.8,3.5){\makebox(0,0){the north pole}}
   \put(1.8,3){\makebox(0,0){$(0,0,1)$}}
   \multiput(1.4,2.1)(0.6,0.4){8}{\line(3,2){0.4}}
   \put(5,4.5){\circle*{0.2}}
   \bezier{40}(4.7,4.3)(4.6,4.45)(4.5,4.6)
   \bezier{40}(4.5,4.6)(4.65,4.7)(4.8,4.8)
   \put(4.4,5.4){\line(2,-3){3}}
   \put(7,1.5){\circle*{0.2}}
   \put(4.5,2.1){\makebox(0,0){$h(q)$}}
   \put(7.6,1.6){\makebox(0,0){$h(p)$}}
   \put(5.4,4.4){\makebox{$h(\tilde{q})$}}
   \put(6,3.1){\makebox{$h(C(\tilde{q}))$}}
   \put(5.9,0.5){\makebox(0,0){$h(C(q))$}}
\end{picture}
\end{center}
\caption{The point $p$ can be reached from $q$.}
\label{figure:belowcq}
\end{figure}
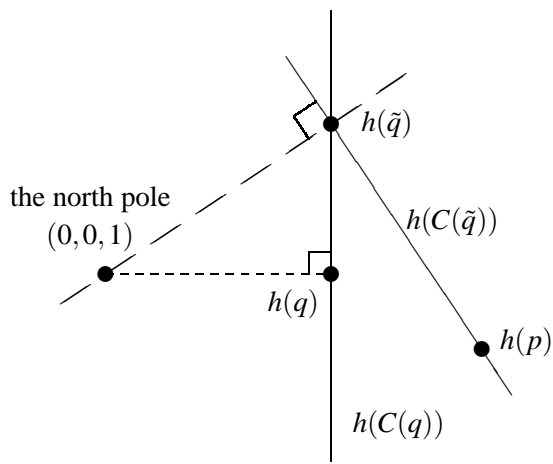

Step 1.
First, we show that, if
$p$ and $q$ are points in the northern
hemisphere and $p$ lies in the
region between $C(q)$ and the
equator, then $p$ can be reached from $q$. In
fact, we show that
there is a point $\tilde{q}$ on $C(q)$ such that $p$
lies on $C(\tilde{q})$.
Therefore we consider the images of $q$ and $p$ in
the plane $H$;
see Figure \ref{figure:belowcq}.
The point $h(p)$ lies in
the half plane bounded by $h(C(q))$ not containing
the north
pole.
Among all points $h(q^\prime)$
on the line $h(C(q))$ we set $\tilde{q}$
to be one of the two points such
that the line trough the north pole and
$h(q^\prime)$
and the line through
$h(q^\prime)$ and $h(p)$ are orthogonal. Then this last
line is the image
of $C(\tilde{q})$, and $C(\tilde{q})$ contains the point $p$.
Hence $p$ can
be reached from $q$. Our first claim is proved.

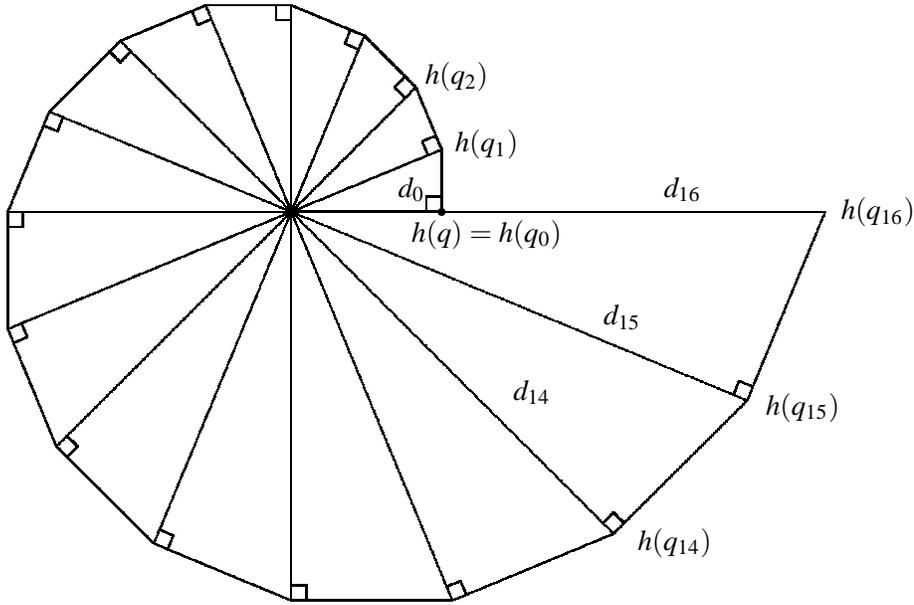
\begin{figure}[htbp]
\setlength{\unitlength}{2cm}
\begin{center}
\begin{picture}(6,5)(-2,-3)
   \thicklines
   \bezier{200}(0,0)(.500, 0)(1.000, 0)
   \thinlines
   \bezier{200}(0,0)(.500, .207)(1.000, .414)
   \bezier{200}(0,0)(.414, .414)(.828, .828)
   \bezier{200}(0,0)(.243, .586)(.485, 1.172)
   \bezier{200}(0,0)(0, .686)(0, 1.373)
   \bezier{200}(0,0)(-.284, .686)(-.569, 1.373)
   \bezier{200}(0,0)(-.569, .569)(-1.137, 1.137)
   \bezier{200}(0,0)(-.804, .333)(-1.608, .666)
   \bezier{200}(0,0)(-.942, 0)(-1.884, 0)
   \bezier{300}(0,0)(-.942, -.390)(-1.884, -.780)
   \bezier{300}(0,0)(-.780, -.780)(-1.561, -1.561)
   \bezier{300}(0,0)(-.457, -1.104)(-.914, -2.207)
   \bezier{300}(0,0)(0, -1.293)(0, -2.586)
   \bezier{300}(0,0)(.536, -1.293)(1.071, -2.586)
   \bezier{300}(0,0)(1.071, -1.071)(2.142, -2.142)
   \bezier{300}(0,0)(1.515, -.627)(3.030, -1.255)
   \bezier{300}(0,0)(1.775, 0)(3.549, 0)

   \bezier{200}(1.000, 0)(1.000, .207)(1.000, .414)
   \bezier{200}(1.000, .414)(.914, .621)(.828, .828)
   \bezier{200}(.828, .828)(.657, 1.000)(.485, 1.172)
   \bezier{200}(.485, 1.172)(.243, 1.272)(0, 1.373)
   \bezier{200}(0, 1.373)(-.284, 1.373)(-.569, 1.373)
   \bezier{200}(-.569, 1.373)(-.853, 1.255)(-1.137, 1.137)
   \bezier{200}(-1.137, 1.137)(-1.373, .902)(-1.608, .666)
   \bezier{200}(-1.608, .666)(-1.746, .333)(-1.884, 0)
   \bezier{200}(-1.884, 0)(-1.884, -.390)(-1.884, -.780)
   \bezier{200}(-1.884, -.780)(-1.722, -1.171)(-1.561,-1.561)
   \bezier{200}(-1.561, -1.561)(-1.238, -1.884)(-.914, -2.207)
   \bezier{200}(-.914, -2.207)(-.457, -2.397)(0, -2.586)
   \bezier{200}(0, -2.586)(.536, -2.586)(1.071, -2.586)
   \bezier{200}(1.071, -2.586)(1.607, -2.364)(2.142, -2.142)
   \bezier{200}(2.142, -2.142)(2.586, -1.699)(3.030, -1.255)
   \bezier{200}(3.030, -1.255)(3.289, -.627)(3.549, 0)

   \put(1,0){\circle*{0.06}}
   \put(1.3,-0.15){\makebox(0,0){$h(q)=h(q_0)$}}
   \put(3.9,0){\makebox(0,0){$h(q_{16})$}}
   \put(0.8,0.15){\makebox(0,0){$d_0$}}
   \put(1.1,0.9){\makebox(0,0){$h(q_2)$}}
   \put(1.3,0.45){\makebox(0,0){$h(q_1)$}}
   \put(2.55,-2.2){\makebox(0,0){$h(q_{14})$}}
   \put(3.4,-1.3){\makebox(0,0){$h(q_{15})$}}
   \put(1.6,-1.2){\makebox(0,0){$d_{14}$}}
   \put(2.2,-0.7){\makebox(0,0){$d_{15}$}}
   \put(2.6,0.15){\makebox(0,0){$d_{16}$}}

   \bezier{40}(.900, 0)(.900, .050)(.900, .100)
   \bezier{40}(.908, .376)(.889, .422)(.870, .468)
   \bezier{40}(.757, .757)(.722, .793)(.686, .828)
   \bezier{40}(.447, 1.080)(.401, 1.099)(.355, 1.118)
   \bezier{40}(0, 1.273)(-.050, 1.273)(-.100, 1.273)
   \bezier{40}(-.531, 1.281)(-.577, 1.262)(-.623, 1.243)
   \bezier{40}(-1.066, 1.066)(-1.102, 1.031)(-1.137, .995)
   \bezier{40}(-1.516, .628)(-1.535, .582)(-1.554, .536)
   \bezier{40}(-1.784, 0)(-1.784, -.050)(-1.784, -.100)
   \bezier{40}(-1.792, -.742)(-1.773, -.788)(-1.754, -.834)
   \bezier{40}(-1.490, -1.490)(-1.455, -1.526)(-1.419, -1.561)
   \bezier{40}(-.876, -2.115)(-.830, -2.134)(-.784, -2.153)
   \bezier{40}(0, -2.486)(.050, -2.486)(.100, -2.486)
   \bezier{40}(1.033, -2.494)(1.079, -2.475)(1.125, -2.456)
   \bezier{40}(2.071, -2.071)(2.107, -2.036)(2.142, -2.000)
   \bezier{40}(2.938, -1.217)(2.957, -1.171)(2.976, -1.125)

   \bezier{40}(1.000, .100)(.950, .100)(.900, .100)
   \bezier{40}(.962, .506)(.916, .487)(.870, .468)
   \bezier{40}(.757, .899)(.722, .864)(.686, .828)
   \bezier{40}(.393, 1.210)(.374, 1.164)(.355, 1.118)
   \bezier{40}(-.100, 1.373)(-.100, 1.323)(-.100, 1.273)
   \bezier{40}(-.661, 1.335)(-.642, 1.289)(-.623, 1.243)
   \bezier{40}(-1.208, 1.066)(-1.173, 1.031)(-1.137, .995)
   \bezier{40}(-1.646, .574)(-1.600, .555)(-1.554, .536)
   \bezier{40}(-1.884, -.100)(-1.834, -.100)(-1.784, -.100)
   \bezier{40}(-1.846, -.872)(-1.800, -.853)(-1.754, -.834)
   \bezier{40}(-1.490, -1.632)(-1.455, -1.597)(-1.419, -1.561)
   \bezier{40}(-.822, -2.245)(-.803, -2.199)(-.784, -2.153)
   \bezier{40}(.100, -2.586)(.100, -2.536)(.100, -2.486)
   \bezier{40}(1.163, -2.548)(1.144, -2.502)(1.125, -2.456)
   \bezier{40}(2.213, -2.071)(2.178, -2.036)(2.142, -2.000)
   \bezier{40}(3.068, -1.163)(3.022, -1.144)(2.976, -1.125)
\end{picture}
\end{center}
\caption{The shell
in the plane $H$ for $n=16$.}
\label{figure:schnecke}
\end{figure}

Step 2.
Fix a point $q$ in
the northern hemisphere.
Starting from $q$ we can wander around the
northern hemisphere along
great circles of the form $C(p)$ for points $p$
in the following way:
for $n\geq 5$ we define a sequence $q_0, q_1, \ldots,
q_n$
by setting $q_0=q$ and by choosing $q_{i+1}$ to be that point on
the
great circle $C(q_i)$ such
that the angle between $h(q_{i+1})$ and $h(q_i)$
is
$2\pi/n$. The image in $H$ of this configuration is a
shell where
$h(q_n)$ is the point furthest away
from the north pole;  see Figure
\ref{figure:schnecke}.
First, we claim that any point $p$ on the unit sphere with
$p_z < {q_n}_z$ can be reached from $q$. Indeed, such a point
corresponds to a point $h(p)$ which is further away from the
north pole than $h(q_n)$.
There is an index $i$
such that $h(p)$ lies in the half plane bounded by  $h(C(q_i))$
and not containing the north pole, hence such that
$p$ lies in the region between $C({q_i})$ and the equator.
Then, as we have already seen, $p$ can be reached from
$q_i$ and hence also from $q$.
Secondly, we claim that $q_n$ approaches
$q$ as $n$ tends to infinity. This is equivalent to
showing that the distance of $h(q_n)$ from $(0,0,1)$
approaches the distance of $h(q)$ from $(0,0,1)$.
Let $d_i$ denote the distance of $h(q_i)$ from $(0,0,1)$
for $i=0,\ldots,n$. Then
$d_i / d_{i+1} = \cos(2\pi/n)$, see Figure
\ref{figure:schnecke}. Hence
$d_n = d_0 \cdot (\cos(2\pi/n))^{-n}$.
That $d_n$ approaches $d_0$ as $n$ tends to infinity
follows immediately from the fact that
$(\cos(2\pi/n))^n$ approaches $1$ as $n$ tends to infinity.
 For
completeness sake\footnote{Actually, this is an exercise in elementary
analysis.} we prove it by
proving the equivalent statement that $\log((\cos(2\pi/n))^n)$ tends to $0$
as $n$ tends to
infinity. Namely, for small $x$ we know the formulae
$\cos(x)=1-x^2/2 + {\cal O}(x^4)$ and
$\log(1+x)=x+{\cal O}(x^2)$.
Hence, for large $n$,
\begin{eqnarray*}
   \log((\cos(2\pi/n))^n) & = &
      n \cdot \log(1-2{\pi^2 \over n^2} + {\cal O}(n^{-4})) \\
   & = & n \cdot ( - 2 {\pi^2 \over n^2} + {\cal O}(n^{-4}))\\
       & = & - {2 \pi^2 \over n} + {\cal O}(n^{-3}) \, .
\end{eqnarray*}
This ends the proof of the geometric lemma.

\appendix
\section*{Appendix B: Proof of a property of the set of consequences
of a theory}

In Section \ref{section:injorder} we introduced
the set $Con({\cal K})$ of consequences of a set ${\cal K}$
of propositions over a set $U$ of {\em simple propositions} and
the logical connectives negation $\phantom{x}^\prime$ and
implication $\rightarrow$.
We mentioned four properties of the operator $Con$.
In this appendix we prove the fourth property:
\[ Con({\cal K}) = \bigcup_{\{X\subseteq {\cal K}, X
\;\mbox{ finite}\}} Con(X) \,.\]

The inclusion
$Con({\cal K}) \supseteq \bigcup_{\{X\subseteq {\cal K}, X
\;\mbox{ finite}\}} Con(X)$ follows directly from
the second property of $Con$,
i.e., from the monotonicity:
if $X \subseteq {\cal K}$, then $Con(X) \subseteq Con ({\cal K})$.
For the other inclusion
we assume that a proposition $A \in Con({\cal K})$
is given. We have to show that there exists a finite
subset $X \subseteq {\cal K}$ such that $A \in Con(X)$.

In order to do this we consider the set ${\cal V}(W(U))$
of all valuations. This set can be identified with the power
set of $U$ and viewed as a topological space with
the product topology of $|U|$ copies of the
discrete topological space $\{0,1\}$. By Tychonoff's
Theorem (see Munkres \cite{munkres-75})
${\cal V}(W(U))$ is a compact topological space.
For an arbitrary proposition $B$
and valuation $t$ the set
$\{t \in {\cal V}(W(U)) \mid t(B)=0\}$
of valuations $t$ with $t(B)=0$
is a compact and open subset of valuations
because the value $t(B)$ depends only on the
finitely many simple propositions occurring in $B$.

Note that our assumption $A \in Con({\cal K})$ is
equivalent to the inclusion
\[ \{t \in {\cal V}(W(U)) \mid t(A)=0\}
\subseteq \bigcup_{B \in {\cal K}}
  \{t \in {\cal V}(W(U)) \mid t(B)=0\}.\]
Since the set on the left-hand side is compact, there exists
a finite subcover of the open cover on the right-hand side,
i.e.\ there exists a finite set $X \subseteq {\cal K}$
with
\[ \{t \in {\cal V}(W(U)) \mid t(A)=0\}
\subseteq \bigcup_{B \in X}
  \{t \in {\cal V}(W(U)) \mid t(B)=0\}.\]
This is equivalent to $A \in Con(X)$ and was to
be shown.

\section*{Acknowledgement}

The authors thank the anonymous referees for their extremely helpful
suggestions and comments leading to a better form of the paper.


 \end{document}